\documentclass{article}
\usepackage[round]{natbib}
\newif\ifcageofig
\cageofigfalse 
\usepackage{shortex}

\usepackage[labelfont=bf,font=footnotesize]{caption} 

\usepackage{enumitem}
\setlist[1]{leftmargin=12pt,itemsep=-3pt,parsep=1ex}

\newlength{\myfigwidth}
\ifcageofig
    \newenvironment{myfig}{\begin{figure}}{\end{figure}}
\else
    \newenvironment{myfig}{\begin{figure}[t]}{\end{figure}}
\fi

\newcommand{\nlayers}{n_\text{layers}}
\newcommand{\nrows}{n_\text{rows}}
\newcommand{\ncols}{n_\text{cols}}
\newcommand{\npps}{n_\text{pps}}
\newcommand{\npixels}{n_\text{pixels}}
\newcommand{\nsensors}{n_\text{sensors}}
\newcommand{\ngrid}{n_\text{grid}}

\newcommand{\nchains}{n_\text{chains}}
\newcommand{\nsuper}{n_\text{super}}
\newcommand{\nwithin}{n_\text{within}}
\newcommand{\heaviside}{\eta}

\title{GPU-accelerated Bayesian inference for block-cave geometry recovery via muon tomography}
\author{
\normalsize Miguel Biron-Lattes\textsuperscript{1}, Patrick Belliveau\textsuperscript{2},
Faezeh Yazdi\textsuperscript{1}, Samopriya Basu\textsuperscript{1}, \\
\normalsize Donald Estep\textsuperscript{1}, Derek Bingham\textsuperscript{1}, 
and Doug Schouten\textsuperscript{2} \\[1em]
\small\textsuperscript{1}Department of Statistics and Actuarial Science, Simon Fraser
University \\[-1pt]
\small\textsuperscript{2}Ideon Technologies Inc.
}
\date{\normalsize\today}

\begin{document}

\maketitle

\begin{abstract}
We describe a Bayesian framework for the inverse problem of
geometry recovery of block caving via muon tomography. We work with a
low dimensional surface-based representation of the geometry of the block cave,
which dramatically reduces the computational requirements of the model while 
allowing realistic geometries. Adopting a Bayesian approach, we define a 
prior distribution on the space of geometries that
favors realistic cave shapes. Pairing this prior with a likelihood based on the
muon tomography forward model, we obtain a posterior distribution over cave 
geometries using Bayes rule. 
We obtain approximate samples from this posterior distribution using 
Markov chain Monte Carlo algorithms running on GPUs, resulting in fast and 
accurate sampling. We test the fidelity of our methodology by applying it to 
a simulated block caving scenario for which the ground truth is known. Results 
show that our method produces sensible geometries that are simultaneously
compatible with the data.
\end{abstract}

\section{Introduction}\label{sec:intro}

\begin{myfig}
\centering
\includegraphics[width=0.5\myfigwidth,trim={70pt 30pt 50pt 0pt},clip]{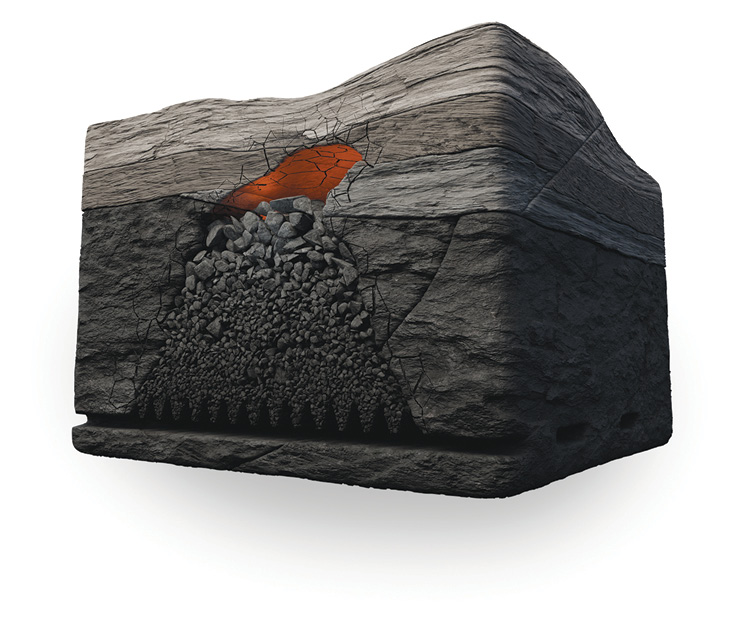}%
\includegraphics[width=0.5\myfigwidth,trim={0pt 0pt 60pt 0pt},clip]{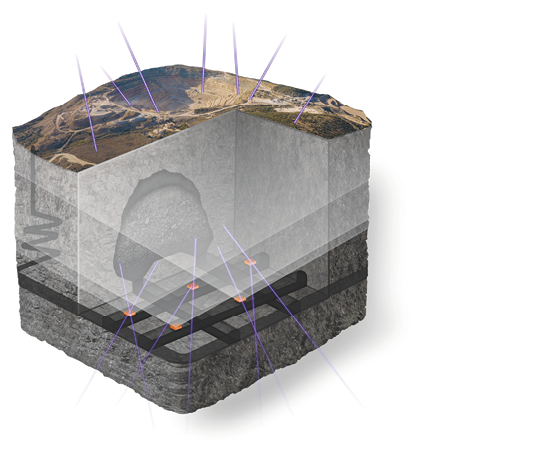}
\vspace{-25pt} 
\caption{Three-dimensional illustration of block caving. 
\textbf{Left:} depiction of the three main units: solid rock, muck pile, and air gap.
Funnels that capture broken ore can be seen below the muck pile.
\textbf{Right:} muons penetrating through the mine, with some intersecting the 
field of view of sensors placed at the bottom of the cave.}
\label{fig:illustration}
\end{myfig}

\emph{Block caving} \citep[see e.g.][]{laubscher1994cave,chitombo2010cave} refers 
to a process that involves undermining an ore body and then allowing it to 
collapse under its own weight. 
The broken ore falls into funnels built underneath the caving 
zone, where it is then extracted (left panel in \cref{fig:illustration}).
This technique provides a way to reach large, lower-grade deposits deep 
underground, achieving high production rates at a fraction of the operating cost 
and environmental impact of conventional surface mining methods.
As such, block cave mining is an effective way to extend the extraction
from an ore deposit when the near-surface ore has been depleted. 

Accurate and timely spatial information of block cave geometry is critical 
for efficient ore recovery and maintaining safety 
\citep{lett2016geotechnical,flores2019major,dawn2019tech}. In particular, 
knowledge of the cave boundaries can help improve safety through monitoring 
hazardous air gap development, enhance the efficiency of the mine, and increase 
accuracy of production forecasts. The problem is thus to determine mine geometry
using mostly indirect observational data. Current methods rely on 
the placement of instrumentation within sparsely distributed drillholes, 
producing time-series data of the locations of trackers. Therefore, this
approach can only provide sparse and localized information about the state
of the cave.

In this paper, we consider using \emph{muon tomography}
\citep{tanaka2013subsurface,schouten2018muon,bonechi2020atmospheric,
bonomi2020applications,schouten2022muon} to determine mine geometry.
Briefly, cosmic-ray muons are charged elementary particles that arise naturally 
from cosmic radiation interacting with the Earth's atmosphere. Particle showers 
of muons bombard the Earth steadily and are attenuated by their interaction with
matter along their trajectory. Muons can penetrate deep into the Earth's crust.
Importantly, their attenuation in matter is proportional to the density of 
material the muon passes through. By measuring the flux of surviving muons using 
detectors---positioned in drillholes and beneath the surface (right panel in 
\cref{fig:illustration})---the average 
density along each ray path within a wide field of view above the 
detector can be determined. 
Each device produces a radiographic image of the rock 
mass above it, and combining the images from several devices yields 
three-dimensional tomographic reconstructions of subsurface density.

Unfortunately, the available data are insufficient to produce a direct image of
the geometry of the cave back, muck pile, or air gap. As a result, most 
of the cave structure is unknown, including the geometry of any possible air 
gap. Realizing the full potential of this technology requires solving the
inverse problem associated with tomography; namely, inverting from the muon
attenuation data to 
the distribution of mass in the cave. Muon tomography-derived density models 
have been successfully deployed to monitor block cave propagation remotely 
\citep{schouten2022muon,schouten2024cosmic}. Still, there remain unsolved 
challenges in the ability to produce a solution that reflects the uncertainty 
in the geometry.

In this work we develop a Bayesian model for the determination 
of block caving geometry via muon tomography. 
Our modeling approach exploits key features of the problem in order to
drastically reduce the dimension of the unknown variables representing the
geometry, enabling fast 
evaluation of the resulting forward model. In turn, this allows us to
perform inference on this model using modern Markov chain Monte Carlo 
(MCMC; see e.g. 
\citealp{geyer1992practical,tierney1994markov,gilks1995markov,brooks2011handbook})
software that leverage the recent increase in availability of Graphic Processing
Units (GPUs) designed to accelerate numerical computations
\citep{phan2019composable}.

To check the fidelity of our proposed methodology, we apply it to a simulated 
block caving scenario for which the ground truth is known. The simulations show
that the MCMC chains converge to the posterior probability distribution on cave
geometries. In particular, the samples obtained produce useful risk assessments
of the existence of an air gap at the top of the cave. This suggests that our 
proposed method constitutes a principled and effective approach for muon
tomography inversion with strong uncertainty quantification. We finish by
discussing potential avenues of research towards improving the models and
samplers.

\section{Methodology}\label{sec:model}

We begin by defining the forward model that relates subsurface density to muon
count data. We then describe how we formulate block-cave imaging with muon
tomography as a geometric inverse problem by partitioning the subsurface into
three units with distinct density distributions. Since we aim to apply MCMC
samplers that utilize gradient information
for proposal generation (see \cref{sec:mcmc}), we focus on a version of the
computer model that is end-to-end differentiable. Finally, we describe the
Bayesian approach to the inverse problem, which involves defining a prior
probability distribution over feasible geometries.

\subsection{Description of the forward model}

In the block-cave imaging problem we are interested in recovering the location
of the interfaces between different sections of the subsurface. 
Specifically, we partition the subsurface into three units: \emph{in-situ}
rock, muck pile, and possibly an air-gap.
While it will never be known exactly in practice, the \emph{in-situ} density will
typically be well characterized prior to caving \citep{laubscher2000block}. 
We can also estimate the expected reduction in density as rock fractures and
becomes part of the muck pile.
Thus, in the following we assume the \emph{in-situ} rock and muck densities
to be approximately known and well separated enough that we can keep them fixed
and infer the interfaces between the muck pile, \emph{in-situ} rock, and possible
air-gap.

\subsubsection{Muon attenuation}\label{sec:forward_model}

\begin{myfig}
\centering
\begin{subfigure}[b]{\myfigwidth}
\includegraphics[width=\textwidth]{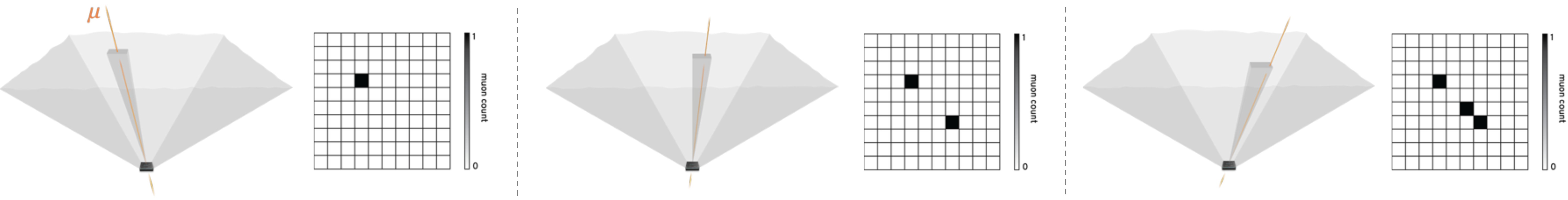}
\caption{Visualization of muon data collection. Each muon that passes through
the detector is recorded. The direction of travel and the position at which the
muon hit the detector are both measured. The radiographic image is set up so
that the centre of the image is straight up. Each pixel is a particular part of
the full field of view, a section of solid angle.}
\label{fig:radiograph-illustration-a}
\end{subfigure}\\[0.4cm]
\begin{subfigure}[b]{\myfigwidth}
\includegraphics[width=\textwidth]{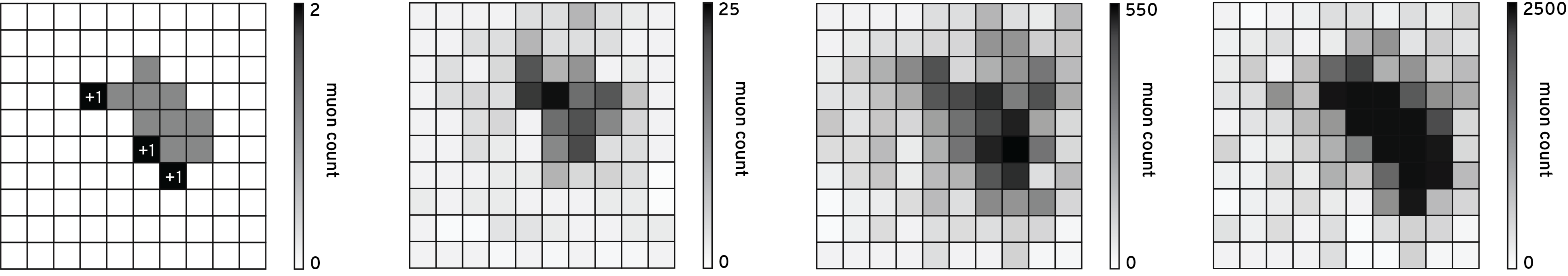}
\caption{Time evolution of a muon counts radiograph. Each time a muon is detected,
we add $1$ to the counter of the corresponding radiograph pixel.}
\label{fig:radiograph-illustration-b}
\end{subfigure}
\caption{Visualization of the muon data collection process for a single sensor
with $\npps=10^2=100$ pixels.}
\label{fig:radiograph-illustration}
\end{myfig}

Muon tomography detectors count the muons passing through them and measure their
trajectories. Each sensor is placed at a known position in the inversion domain
$\mathcal{D}$---which for simplicity we assume has the shape of a box---with
a given field of view specified by a solid angle $\Omega$. The muon trajectories
captured by the detector are binned by solid angle into $\npps$ \emph{pixels}
per sensor, each with a specific portion $\Omega_p$ of solid angle so that 
$\{\Omega_p\}_{p=1}^{\npps}$ defines a partition of $\Omega$.
As \cref{fig:radiograph-illustration} shows, this data can be represented in a
2D image called a \emph{radiograph}. Thus, the data used in our inference problem 
corresponds to the collection of observed counts across all pixels in every
available detector. Interested readers may see \citet{schouten2024cosmic} for a
more technical description of the types of sensors considered in our models.

Given a density field $\rho:\mathcal{D}\to[0,\infty)$, the arrival of muons 
at each pixel is modeled as a Poisson process conditionally independent of other pixels.
Specifically, for any exposure time $\Delta t>0$, the number of muons detected at
the $p$-th pixel is assumed to follow a Poisson distribution with mean
\[\label{eq:expected_counts}
\lambda_p(\rho) := \Delta t \int_{\Omega_p} \mathcal{I}\left( \mathcal{O}(\rho,\hat{n}) \right) \varepsilon(\hat{n}) \d\theta,
\]
where $\Omega_p$ is the solid angle associated with the $p$-th pixel; 
$\mathcal{I}$ is the muon intensity function;
$\hat{n}=\hat{n}(\theta)$ is the normal vector associated to a given
$\theta\in\Omega_p$;
$\varepsilon(\hat{n})$ is the efficiency of the detector as a function of the 
direction $\hat{n}$, and
\[
\mathcal{O}(\rho,\hat{n}) := \int_{\varrho(\hat{n})}\rho(x(u),y(u),z(u))\mathrm{d}u
\]
is the opacity along the ray $\varrho(\hat{n})$ that starts at the sensor and 
has direction $\hat{n}$.
$\mathcal{I}$ is determined by the muon source, and $\varepsilon$ is related to
how well the device detects muons incident from the specified direction.
We refer the reader to \citet{schouten2022muon} for a detailed description of these
functions.

Practical implementations of \cref{eq:expected_counts} employ 
numerical approximations to the integrals. One approximation is given by piece-wise 
constant fields defined on a grid. Specifically, we discretize the 
domain $\mathcal{D}$ into a regular three-dimensional grid with steps 
$(\Delta x, \Delta y, \Delta z)$ and sizes $(n_x,n_y,n_z)$---implying a total of
$\ngrid=n_xn_yn_z$ grid elements or \emph{voxels}.

\begin{myfig}
	\centering
	\includegraphics[width=\myfigwidth]{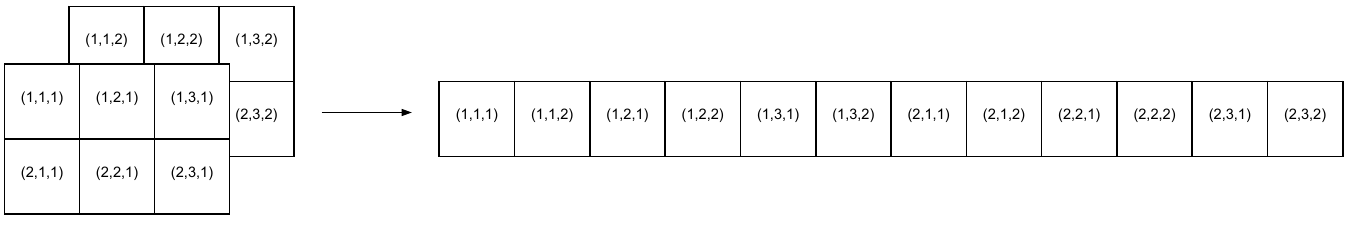}
	\caption{An example of reshaping a three-dimensional array of size $(2,3,2)$
		into a vector of length $12$, using the convention in which the right-most index 
		varies the fastest.}
	\label{fig:array_indexing}
\end{myfig}

Using the density-grid we represent piecewise constant density fields
as three-dimensional arrays $R$. It is helpful to view $R$
as a vector of $\ngrid$ ``stacked'' voxels (see \cref{fig:array_indexing}). 
Any linear indexing of 
the array is theoretically valid, but picking the memory layout convention 
used by the language in which the model is implemented is most efficient as 
it makes the translation between the two types of indices computationally free.
We overload the notation and let $\lambda_p:\reals^{\ngrid}\to\reals^+$ 
be the restriction of \cref{eq:expected_counts} to piece-wise constant density 
fields defined on the grid.

\subsubsection{Speeding-up model evaluation through linearization}

It is known that the uncertainty quantification benefits of using MCMC 
for sampling the posterior distribution of an inverse problem 
can come at a high computational cost, due to the number of model evaluations 
the method requires \citep{mosegaard1995monte}. While several approaches 
exist to tackle this issue \citep{higdon2003markov, christen2005markov}, 
in this work we opt for a simpler alternative.
Leveraging the fact that $R\mapsto \lambda_p(R)$ is
differentiable, we work with a linear approximation around a
reference density array $R_0$. Consider the first-order Taylor expansion
\[\label{eq:expected_counts_linearized}
\lambda_p(R) = \lambda_p(R_0) + \frac{\partial \lambda_p}{\partial R}(R_0)^\top(R-R_0) + O(\|R-R_0\|^2),
\]
where $\frac{\partial \lambda_p}{\partial R}(R_0)$ is the gradient of the expected
counts evaluated at the reference grid. Stacking 
\cref{eq:expected_counts_linearized} across all $\npps$ pixels in all $\nsensors$ 
sensors and dropping higher order terms yields a linear approximation to the 
forward model
\[\label{eq:expected_counts_linearized_approx}
\hat\lambda(R) := \lambda(R_0) + G(R-R_0).
\]
Here, both $\hat\lambda(R)$ and $\lambda(R_0)$ are stacked vectors of 
$\npixels:=\nsensors\cdot\npps$ expected muon counts, while $G$ is the
$\npixels \times \ngrid$ \emph{sensitivity} matrix built by stacking the 
gradients at every pixel. Note that $G$ is sparse, as only a handful 
of grid elements interact with the field of view of a given pixel. Therefore, 
using sparse linear algebra routines to compute 
\cref{eq:expected_counts_linearized_approx} results in a fast approximation to
the expected muon counts.

We stress that the linearization of $R\mapsto \lambda_p(R)$ does not make the
full inverse problem linear, as we still are missing a component model linking
cave geometry to a density array. The next section shows that this link is, 
in fact, highly non-linear.

\subsubsection{Low-dimensional representation of the cave geometry}

\begin{myfig}
\centering
\includegraphics[width=0.9\myfigwidth,trim={22pt 5pt 6pt 11pt},clip]{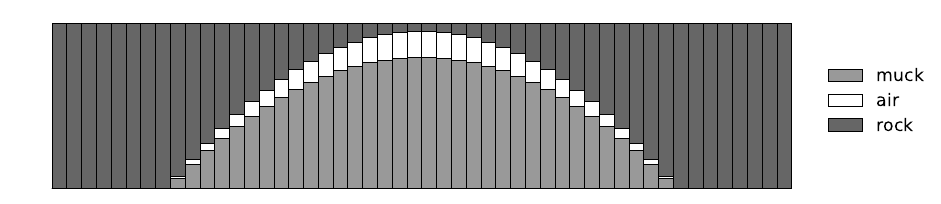}
\caption{A two-dimensional visualization of a layer model with $3$ distinct units
representing rock, muck pile, and possible air gap.}
\label{fig:example_layermodel}
\end{myfig}

To represent the geometry of a block cave, we choose a surface-based approach,
in which the subsurface is represented by the location of the interfaces 
between the distinct units of which it is comprised. 
Referring to \cref{fig:illustration}, we show three distinct units: solid rock, 
muck pile, and air gap. When the material properties of each unit are 
known to a high degree, the uncertainty lies in their shape and location. 
In turn, these can be specified by the location of their interfaces. This approach 
dramatically reduces the number of variables needed to specify a geometry, as it
transforms a three-dimensional problem---the specification of the mass density at
every point in the inversion domain---into the two-dimensional issue of finding
vertical limits for the interfaces between geological units.

Specifically, we use \emph{layer models} for geological surfaces,
consisting of a collection of layers defined on a fixed grid with specified 
thickness. The layers are arranged sequentially
to represent a three-dimensional object. \cref{fig:example_layermodel} shows a 
two-dimensional visualization of a three layer model.
This family of models is prevalent in the Earth sciences to represent portions
of the subsurface that are known to consist of layered sedimentary strata
\citep[see e.g.][]{webber1990framework,turner2006challenges,enemark2022influence}.

We discretize the $x$-$y$ plane of $\mathcal{D}$ into a grid of size 
$\nrows\times \ncols$, and divide the box vertically into
$\nlayers$ layers. This 2D grid need not necessarily match the first two 
coordinates of the 3D density grid described in \cref{sec:forward_model}.
We only require that the two grids be aligned, and that there exist a map 
$(i,j)\mapsto(x(i),y(j))$ taking the first two density-grid indices
into layer-grid indices.
We let $H_{\ell,x,y}\geq 0$ be the relative height with respect to the
floor of $\mathcal{D}$ of the $\ell$-th layer, $\ell\in\{1,\dots,\nlayers\}$, 
at the grid point $(x,y)$.

\begin{myfig}
\centering
\includegraphics[width=0.25\myfigwidth,trim={165pt 10pt 15pt 15pt},clip]{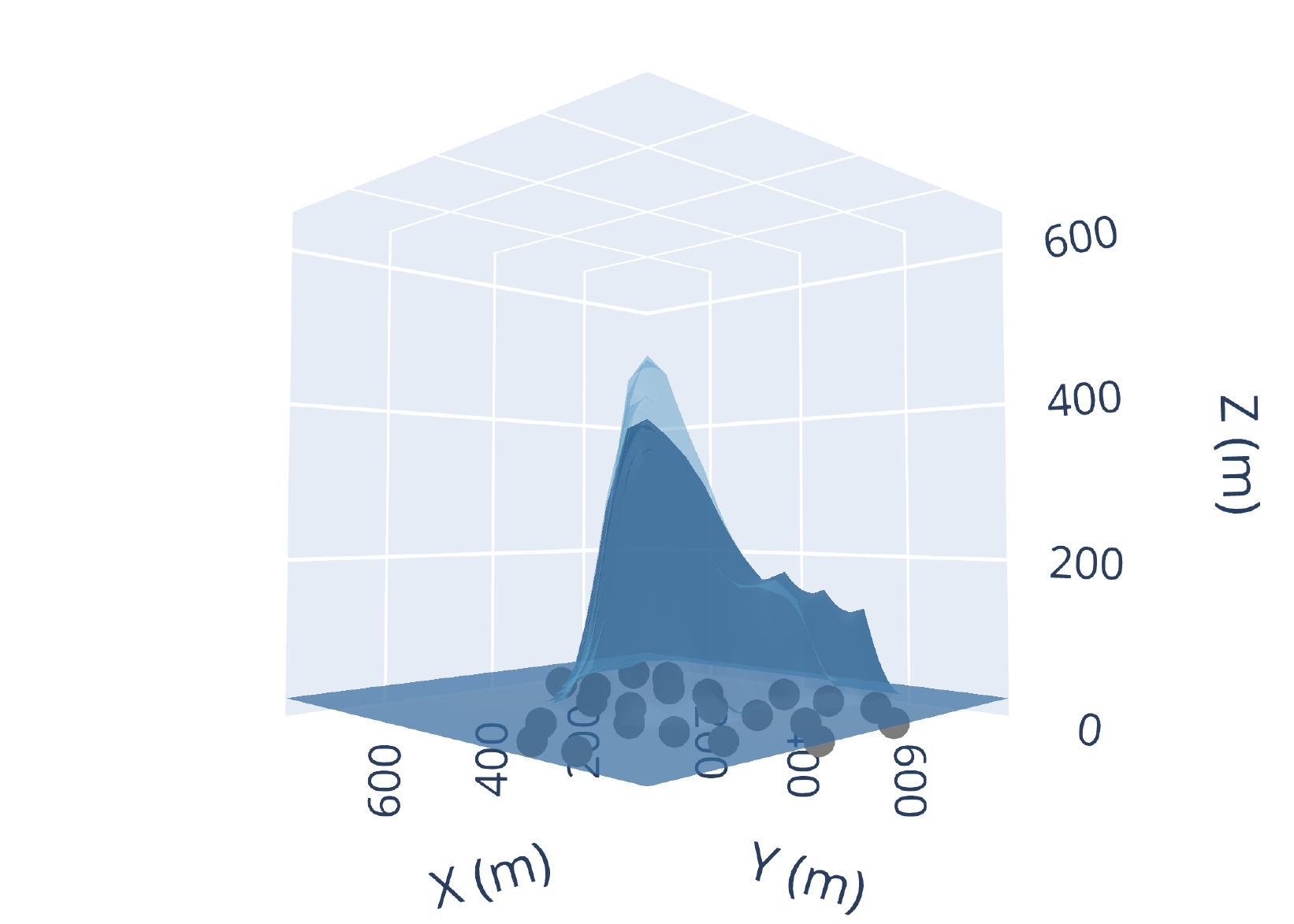}%
\includegraphics[width=0.25\myfigwidth,trim={165pt 10pt 15pt 15pt},clip]{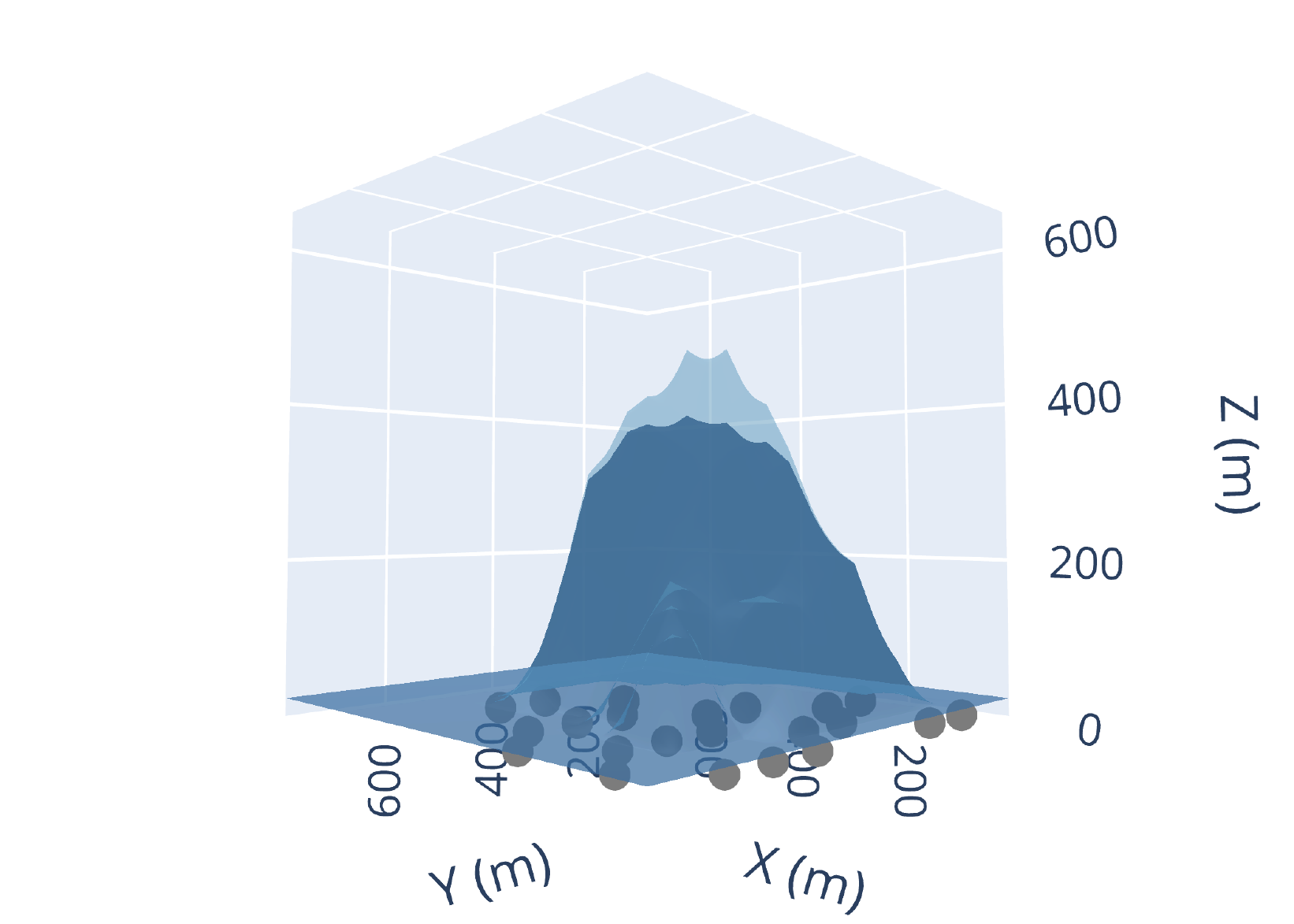}%
\includegraphics[width=0.25\myfigwidth,trim={120pt 10pt 60pt 5pt},clip]{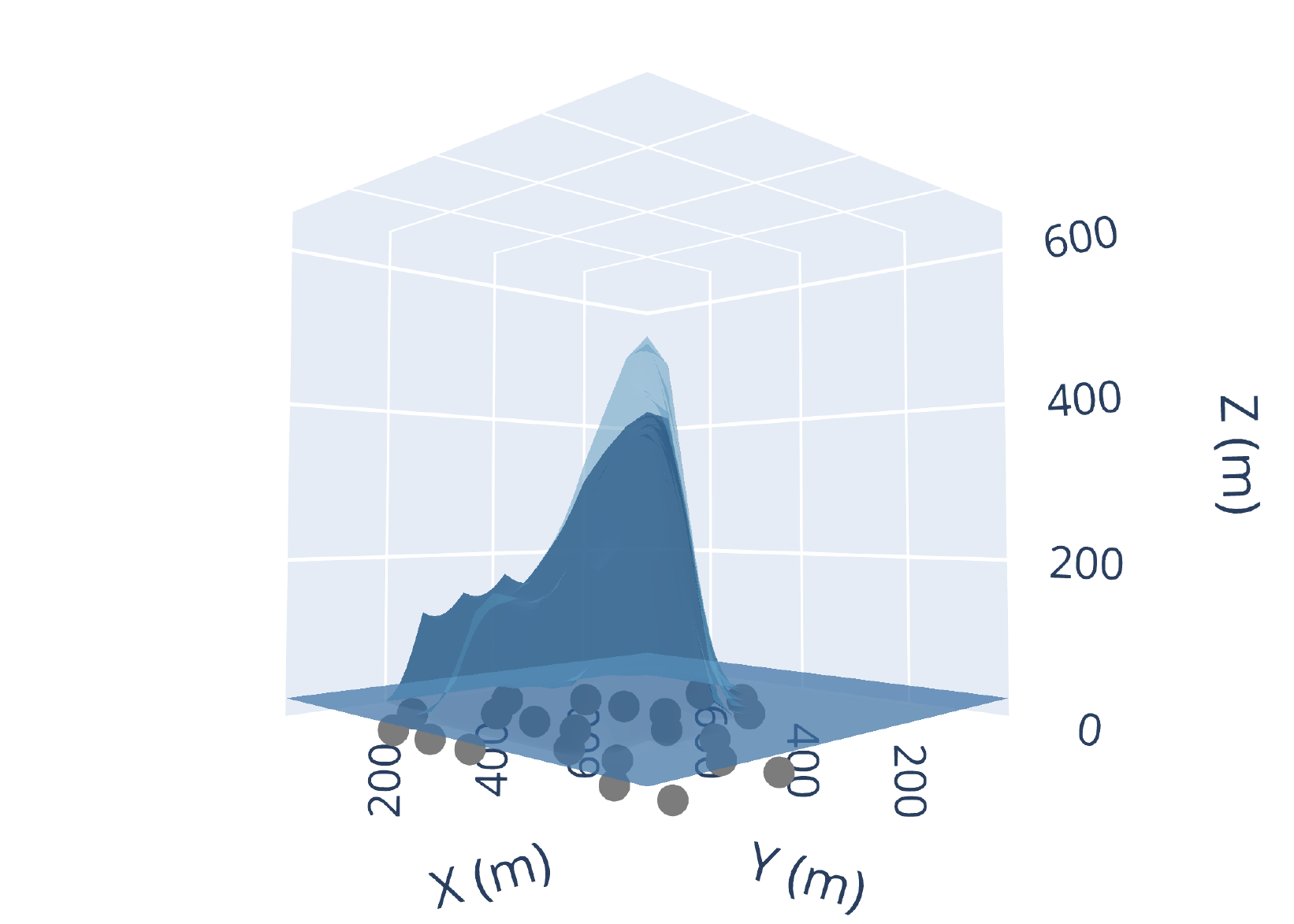}%
\includegraphics[width=0.25\myfigwidth,trim={15pt 10pt 165pt 15pt},clip]{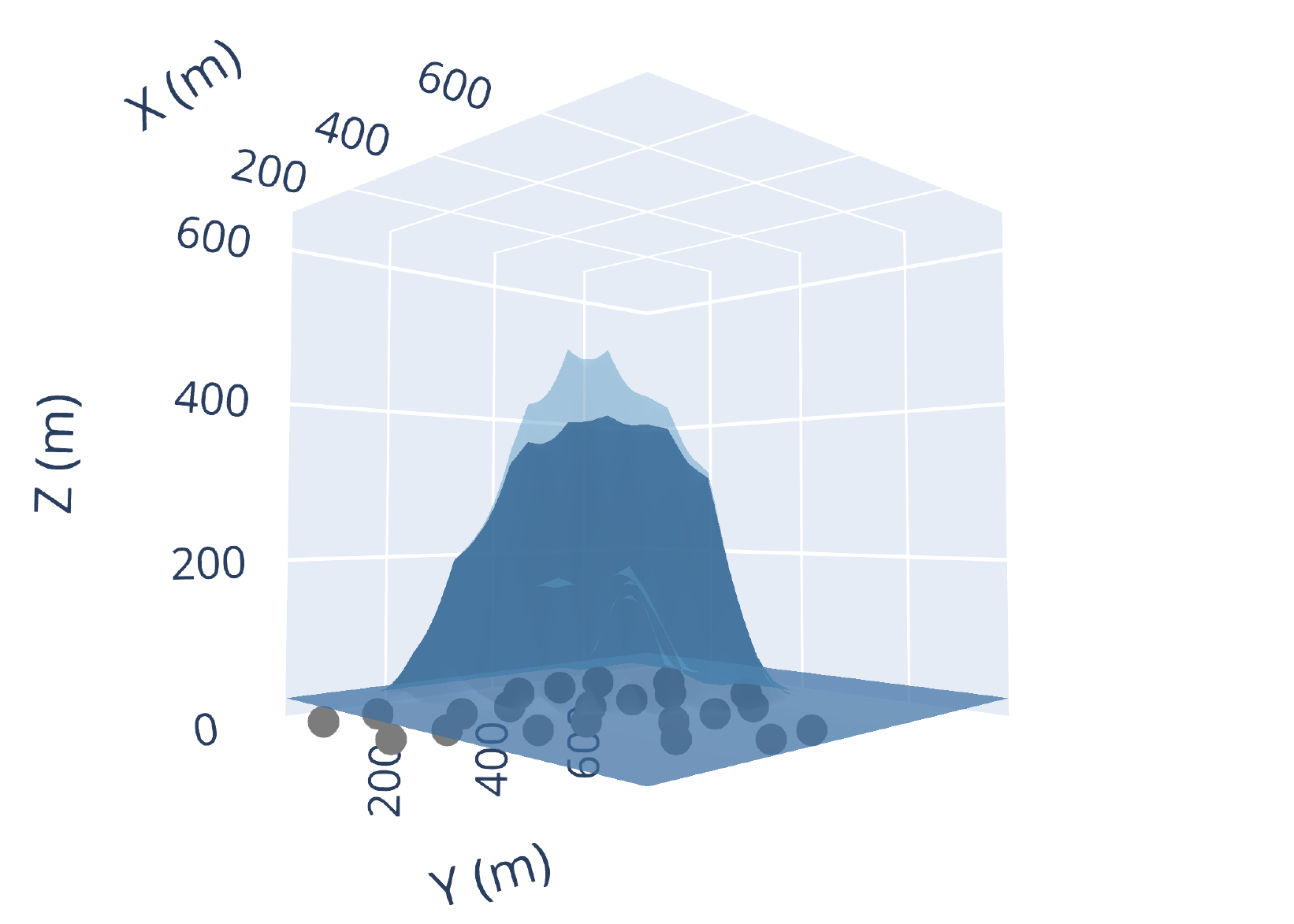}
\caption{Four perspectives of the true layers of the simulated example dataset,
Successive plots from left to right correspond to $90^\circ$ clockwise rotations
about the $z$ axis.
Layers have been plotted with transparency in order to show the non-trivial 
air gap at the top of the formation. Gray markers indicate the location of 
$25$ muon detectors below the cave.}
\label{fig:true_layers}
\end{myfig}

In block caving applications, we represent the geometry using $3$ layers 
corresponding to the interfaces between 1) muck pile and air gap; 2) 
air gap and solid rock, and 3) solid rock and the Earth's surface.
The goal is to perform inference on the thicknesses of the first two layers, 
since the location of the Earth's surface is known.
\cref{fig:true_layers} shows the layer model (without the top of the terrain to 
simplify the visualization) corresponding to a realistic simulated cave model 
provided by a block-caving expert \citep{diering2023deepcove}.
Note that the first layer is flat except where the muck pile is located. Similarly,
the thickness of the air layer is zero everywhere except where the air gap lives.
These zero-thickness regions allow layer models to represent realistic geometries 
while still fundamentally remaining in two-dimensional space---thereby keeping 
computational costs in check. In practice, we enforce
strictly positive thicknesses to avoid numerical issues. This restriction has no
practical effect, as thicknesses can be as close to $0$ as necessary.

\subsubsection{Smoothly mapping layer heights into density grids}

The final step in building the forward model approximation is translating a 
layer model into a density grid. To this end, note that
a voxel $(i,j,k)$ in the grid is below the upper limit of the $\ell$-th layer if
$k\Delta z < H_{\ell,x(i),y(j)}$. It follows that a point in the grid
is inside the $\ell$-th layer if the previous condition holds and also 
$k\Delta z > H_{\ell-1,x(i),y(j)}$ (recall that $H_{0,x,y}=0$).
Since a layer model assumes known densities within
each unit---which we can view as fixed arrays
$\{R^{(\ell)}\}_{\ell=1}^{\nlayers}$---the overall density array $R$ associated
with a given layer model $H$ can be expressed as
\[\label{eq:def_LM_density_grid}
R_{i,j,k}(H) = \sum_{\ell=1}^{\nlayers} [B_{i,j,k}^{(\ell)}(H) - B_{i,j,k}^{(\ell-1)}(H)] R^{(\ell)}_{i,j,k},
\]
where
\[\label{eq:def_below_interface_indicator}
B_{i,j,k}^{(\ell)}(H) := \heaviside(H_{\ell,x(i),y(j)}-k\Delta z),
\]
and $\heaviside:\reals \to [0,1]$ is the Heaviside function, defined as
\[
\heaviside(u) := 
\begin{cases}
0 & u< 0 \\
0.5 & u=0 \\
1 & \text{o.w.}
\end{cases}.
\]
Note that $B_{i,j,k}^{(\ell)}(H)$ plays the role of an indicator function of the 
$(i,j,k)$ voxel being below the upper limit of the $\ell$-th layer.

\begin{myfig}
\centering
\includegraphics[width=0.9\myfigwidth]{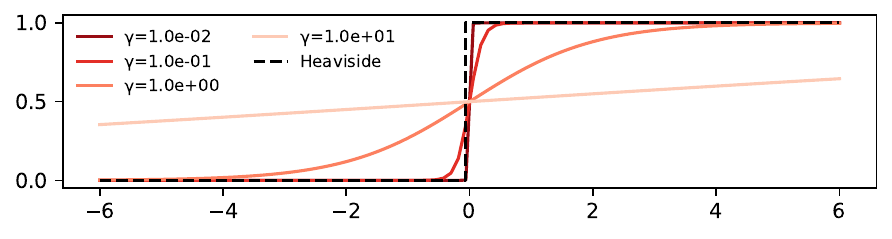}
\caption{Visualization of the approximation of the Heaviside function via
scaled sigmoid functions, for different values of the scale parameter $\gamma$.}
\label{fig:heaviside_approx}
\end{myfig}

As we mentioned at the beginning of \cref{sec:model}, we aim for a 
differentiable forward model that will allow us to use MCMC algorithms that
make use of gradient information. The presence of the Heaviside function breaks
this property. Therefore, we replace the Heaviside with a smooth approximation. 
We employ the scaled sigmoid (or logistic) function
\[\label{eq:def_sigmoid}
\varsigma_\gamma(u) := \frac{1}{1+\exp(-u/\gamma)},
\]
for some $\gamma>0$. \cref{fig:heaviside_approx} shows a visualization of the
behavior of the function for different $\gamma$ values. It is easy to check that 
$\varsigma_\gamma \to \heaviside$ as $\gamma\to 0$. 

Choosing a smoother sigmoid increases the computational efficiency
at a cost of ``smearing'' the boundary between material types, increasing uncertainty
in the result.
Experimentally, we find that values of $\gamma$ around $1$ strike an acceptable 
balance between the two competing effects. Hence, we set $\gamma=1$, and simplify
the notation by writing $\varsigma=\varsigma_1$.

Replacing the Heaviside by the sigmoid in the summand of 
\cref{eq:def_LM_density_grid} yields
\[\label{eq:smooth_layer_indicators}
D_{i,j,k}^{(\ell)}(H) := \varsigma(H_{\ell,x(i),y(j)} - k\Delta z) - \varsigma(H_{\ell-1,x(i),y(j)} - k\Delta z).
\]
Note that $D_{i,j,k}^{(\ell)}(H)\in[0,1)$, because the first term always dominates 
the second term, since layers are increasing in height. Plugging these smooth indicators
into \cref{eq:def_LM_density_grid} gives
\[\label{eq:def_LM_density_grid_smooth_bad}
\sum_{\ell=1}^{\nlayers} D_{i,j,k}^{(\ell)}(H) R^{(\ell)}_{i,j,k}
\]
Unfortunately,
$\sum_{\ell=1}^{\nlayers} D_{i,j,k}^{(\ell)}(H)$ is not guaranteed to be exactly
$1$, as is the case if we work with the Heaviside function. In other words, the 
collection of smooth indicators
defined by \cref{eq:smooth_layer_indicators} is not a partition of unity. Thus,
\cref{eq:def_LM_density_grid_smooth_bad} may produce density values that lie
outside of the range specified by the known density fields
$\{R^{(\ell)}_{i,j,k}\}_{\ell=1}^{\nlayers}$, and therefore are not physically
meaningful. We correct \cref{eq:def_LM_density_grid_smooth_bad} using the
normalization
\[\label{eq:def_LM_density_grid_smooth}
\hat R_{i,j,k}(H) := \sum_{\ell=1}^{\nlayers} W_{i,j,k}^{(\ell)}(H) R^{(\ell)}_{i,j,k}, \qquad W_{i,j,k}^{(\ell)}(H) := \frac{D_{i,j,k}^{(\ell)}(H)}{\sum_{\ell'=1}^{\nlayers} D_{i,j,k}^{(\ell')}(H)},
\]
where the weights $W_{i,j,k}^{(\ell)}(H)$ now define a partition of unity.
Substituting \cref{eq:def_LM_density_grid_smooth} into 
\cref{eq:expected_counts_linearized_approx} then yields a smooth approximation 
$\lambda(\hat R(H))$ to the expected muon count function.

\subsection{Bayesian calibration}

The preceding sections described the forward model of muon tomography, where
a known geometry is used to compute the expected number of 
muons that would be measured by a collection of detectors. We now turn our
attention to the inverse problem, where the goal is to recover the 
geometry given muon counts $\{C_p\}_{p=1}^{\npixels}$ at every pixel. 
As mentioned in \cref{sec:forward_model}, these data are inherently stochastic
and, conditional on the layer heights $H$, follow a Poisson distribution
\[\label{eq:def_likelihood}
C_p|H \overset{\text{indep}}{\sim} \text{Poisson}(\lambda_p(\hat R(H))).
\]

In this work, we use a Bayesian approach for inferring layer heights. In this 
framework \citep[see e.g.][]{mosegaard1995monte,kennedy2001bayesian,
lee2002markov,kaipio2005statistical}, \cref{eq:def_likelihood} corresponds to 
the \emph{likelihood}, defining a probability
distribution of the data conditional on the unknown parameters. The forward model 
is an intrinsic part of the likelihood, as it gives the ``typical'' data that is 
expected for a specific realization of the parameters. In the context of this paper,
the forward model defines the expected muon counts for a given geometry.
The second component in the Bayesian formulation corresponds to the 
\emph{prior} distribution, which encapsulates \emph{a priori} knowledge 
about the unknown parameters. The following section describes a novel prior 
probability distribution for layer models.

After gathering field observations, we update our knowledge about the unknown
parameters using Bayes' theorem to obtain a \emph{posterior}
probability distribution over the unknowns. Conceptually, Bayes' theorem reads
\[
\text{posterior} = \frac{\text{prior}\times \text{likelihood}}{\text{marginal likelihood}}.
\]
The \emph{marginal likelihood} (also known as \emph{evidence}) is a 
normalization constant that ensures the posterior is a probability measure.
Although hard to compute in general \citep{friel2012estimating,
fourment2019dubious}, in many situations---including this work---MCMC can be 
used even if this quantity is unknown, as these algorithms typically only 
require computing ratios of posterior density \citep{tierney1994markov}. We 
will discuss the algorithms we employ to sample from the posterior distribution
in \cref{sec:mcmc}.

\subsubsection{Constructing a prior for layer models}\label{sec:CAR_prior}

They key of the Bayesian approach is choosing an adequate prior.
In this section we focus on defining a prior distribution for layer heights.
We begin with a model that, at grid location $(x,y)$, assigns the height of 
the base layer $H_{1,x,y}$ by sampling uniformly from the available space 
$[0,H_{\nlayers,x,y}]$. By sampling iteratively conditional on the previous layers, we 
choose the height of the $\ell$-th layer uniformly from the remaining space 
$[H_{\ell-1,x,y},H_{\nlayers,x,y}]$. Mathematically,
\[\label{eq:indep_columns}
H_{1,x,y} &\overset{\text{indep}}{\sim} \Unif(0,H_{\nlayers,x,y}) \\
H_{\ell,x,y}|H_{\ell-1,x,y} &\overset{\text{indep}}{\sim} \Unif(H_{\ell-1,x,y}, H_{\nlayers,x,y}), \quad \ell \in \{2,\dots,\nlayers-1\},
\]
where $\Unif(a,b)$ is the uniform distribution on the interval $[a,b]$. 
Note that we can use any distribution with support on a finite interval.
We use the uniform prior as a non-informative choice since, \emph{a priori},
we do not know where the boundaries of these geological units lay.

It is clear that the model in \cref{eq:indep_columns} satisfies the height 
constraints: layers are placed sequentially in order from bottom to top, and they 
all reside inside the enclosing boundaries. However, the model has an important
drawback, since it essentially defines a collection of \emph{independent columns} 
(if $(x_1,y_1)\neq(x_2,y_2)$, the random vectors $H_{\cdot,x_1,y_1}$ and 
$H_{\cdot,x_2,y_2}$ are independent). This allows for very rough surfaces that
are not physically possible, which is especially problematic when the data are
scarce and weakly informative.

One way to introduce spatial structure between columns to \cref{eq:indep_columns} 
starts by noting that it can be re-written as
\[
u_{\ell,x,y} &\distiid \Unif(0,1) \\
H_{1,x,y} &= (1-u_{1,x,y})\cdot0 + u_{1,x,y}H_{\nlayers,x,y} \\
H_{\ell,x,y} &= (1-u_{\ell,x,y})H_{\ell-1,x,y} + u_{\ell,x,y} H_{\nlayers,x,y}, \quad \ell\in \{2,\dots,\nlayers\}.
\]
To allow for within-layer dependence between the heights at different
grid points, we substitute the \iid collection $\{u_{\ell,x,y}\}_{x,y}$ with another 
$\{\check u_{\ell,x,y}\}_{x,y}$ that maintains the marginal $\Unif(0,1)$ distributions 
but has non-trivial dependence. To this end, \emph{copulas} \citep{joe2014dependence} 
provide a framework for modeling dependence among collections of random 
variables while maintaining their marginal distributions. Perhaps the simplest 
copula is the Gaussian. To obtain the collection of dependent 
uniform variates, we first simulate a multivariate normal $\Norm(0,\Sigma)$ in 
$\reals^{\nrows\cdot\ncols}$ with prescribed covariance matrix $\Sigma$,
and then apply an element-wise cumulative distribution function (c.d.f.)
transformation to obtain correlated $\Unif(0,1)$ random variables 
\citep[see e.g.][Example 5.9]{owen2013monte}.

Thus, we have translated the problem of drawing spatially dependent uniform random
variables to sampling from a multivariate normal distribution. It
remains to define an appropriate correlation structure for the layers.
For layer models, the graph defined by the nearest neighbors to every 
grid element provides a natural undirected graph structure.
To this end, conditional autoregressive processes
(CAR, \citealp{besag1974spatial}; see also \citealp{rue2005gaussian}, Section 2.2) 
provide a convenient parametrization of multivariate normal distributions on arbitrary 
graphs $G=(V,E)$, where $V$ is a set of vertices and $E$ a set of edges, assuming the 
graphs contains no self-loops (i.e., $(i,i)\notin E$ for any $i\in V$). CAR processes
are specified from the \emph{full conditional distributions}: if $x\in\reals^{|V|}$---%
with $|V|$ the number of nodes in the graph---is a sample from a CAR process, then 
$x$ follows a multivariate normal distribution satisfying
\[\label{eq:CAR_full_cond_moments}
\E[x_i|x_{-i}] = - \sum_{j\in V:j\sim i} \phi_{i,j}x_j, \qquad \Var(x_i|x_{-i}) = \kappa_i^{-1},
\]
where $x_{-i} =\{x_j: j\neq i\}$; $i\sim j$ means $(i,j)\in E$; 
$\phi$ is a matrix of coefficients, and $\kappa$ is
a vector of conditional inverse variances or \emph{precisions}. Thus, we see that 
a CAR process is specified through the influence that neighborhoods have on vertices.
A pair $(\kappa,\phi)$ yields a valid CAR process specification if the 
\emph{precision matrix}
\[
Q = \diag(\kappa)(I + \phi)
\]
is positive definite, where $\diag(v)$ is the diagonal matrix with diagonal equal
to the vector $v$. In this case, $x\sim \Norm(0,\Sigma)$, where $\Sigma=Q^{-1}$ is
the covariance matrix.

We define the \emph{adjacency matrix} $A$ of the graph $(V,E)$, that has entries
\[
A_{i,j} = 
\begin{cases}
1 & i\sim j\\
0 & \text{otherwise}.
\end{cases}
\]
A simple non-trivial valid CAR process defined on the graph
uses \citep[Section 3]{gelfand2003proper}
\[
\kappa = A\mathbf{1}, \qquad \phi = -r\diag(\kappa^{-1})A,
\]
where $\mathbf{1}=(1,\dots,1)\in\reals^{|V|}$, and $r\in[0,1)$ is a spatial
dependence parameter. Using these expressions in \cref{eq:CAR_full_cond_moments},
we see that this CAR process establishes positive correlation between nodes and 
its neighbors, while it forces nodes with larger neighborhoods to have lower variance. 
Specifically, the nodes are independent when $r=0$, whereas they become increasingly 
more correlated as $r\to 1$. The precision matrix in this case simplifies to
\[\label{eq:def_CAR_precision}
Q = \diag(A\mathbf{1}) - rA,
\]
which is positive-definite for $r\in[0,1)$. For many graphs of interest, $A$ is 
a sparse matrix, since every node is connected to only a handful of neighbors, 
which in turn means that $Q$ is sparse. As we shall soon see, this can be 
computationally advantageous \citep{rue2001fast}.

\begin{myfig}
\centering
\includegraphics[width=0.35\myfigwidth]{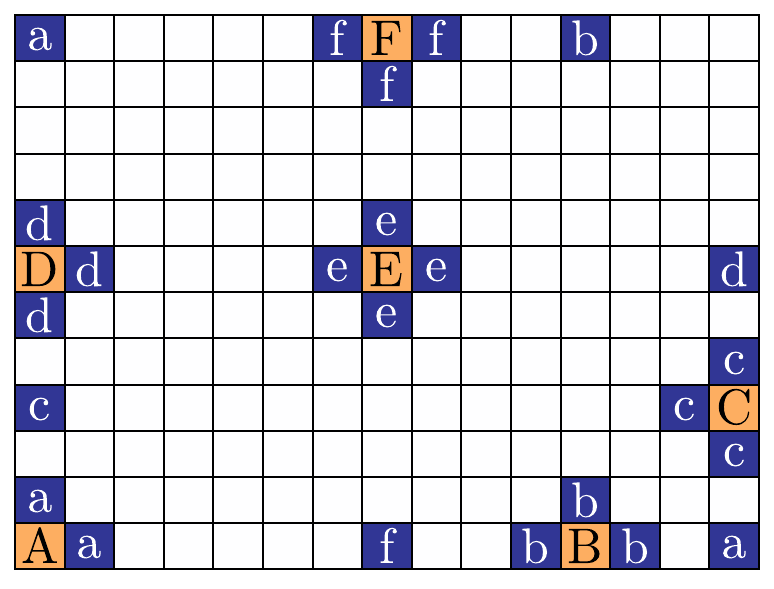}
\caption{Visualization of the neighbors (blue) of selected nodes (yellow) in 
the nearest-neighbor graph of a $12\times15$ grid assuming a periodic domain.
We denote the different selected notes with capital letters, whereas their 
respective neighbors are marked with the corresponding lower-case symbol. 
Notice how the neighborhoods of some nodes wrap around the edges.}
\label{fig:neighbors}
\end{myfig}

Specifying a member of this family of CAR processes thus requires selecting a 
specific graph structure and a scalar $r\in[0,1)$. As we said before, we use the
nearest-neighbor graph induced by the layer grid. We further impose a 
periodic boundary condition, as illustrated in \cref{fig:neighbors}. Doing so
has the effect of making all the node neighborhoods of the same size 
($4$), which in turn means that all the coordinates of the CAR process have the 
same marginal distribution.

We treat the spatial correlation as a hierarchical parameter to be learned from data. 
We assume \emph{a priori} independent per-layer spatial dependency parameters 
$\{r_\ell\}_{\ell=1}^{\nlayers}$ distributed as
\[
r_\ell \overset{\text{\iid}}{\sim} \Unif(0,1).
\]
We choose this distribution because it is non-informative, as in general we do not
have \emph{a priori} knowledge of the strength of spatial association. However,
more informative distributions supported on $(0,1)$---such as specific choices
of the Beta distribution---may be used whenever such knowledge is available.

It remains to specify a procedure to simulate from the CAR process. We start from a 
fundamental property of multivariate normal
distributions. If $z\sim\Norm(0,I)$ is a vector of \iid standard normal variables,
and if there is a matrix $U$ such that $\Sigma=UU^T$, then $x=Uz$ 
follows a $\Norm(0,\Sigma)$ distribution. To leverage this approach, let $L$ be a 
Cholesky factorization of the precision matrix in \cref{eq:def_CAR_precision}; i.e., 
a lower-triangular matrix such that $Q=LL^T$. As we mentioned before, the sparseness
of $Q$ may make this procedure more efficient if sparse linear algebra routines
are available. In any case, consider the identity
\[\label{eq:cov_mat_from_prec_chol}
\Sigma = Q^{-1} = (LL^\top)^{-1} = (L^\top)^{-1}L^{-1} = (L^{-1})^\top L^{-1} = UU^\top,
\]
where $U := (L^{-1})^\top$. It follows that if $z \sim \Norm(0,I)$, 
$x := Uz$ is a sample from $\Norm(0,\Sigma)$ and therefore from the CAR process.
Note that since $L$ is lower-triangular, we can efficiently obtain $x$ by
solving the system $L^\top x=z$ via back-substitution.

With the CAR sample available, we obtain a collection of correlated
$\Unif(0,1)$ random variables via the c.d.f. transform
\[\label{eq:CAR_cdf_transform}
\check u_i := \Phi\left(\frac{x_i}{\sqrt{\Sigma_{i,i}}}\right) 
= \Phi\left(\frac{x_i}{\sqrt{\Sigma_{1,1}}}\right) 
\]
where $\Phi:\reals\to(0,1)$ is the c.d.f. of the standard normal distribution.
The last equality uses the fact that the diagonal of $\Sigma$ is constant, due 
to the symmetry induced by the periodic boundary of the CAR. Moreover, pre and 
post-multiplying \cref{eq:cov_mat_from_prec_chol} by the first vector in the
standard basis $e_1=(1,0,\dots,0)$ yields
\[
\Sigma_{1,1} = e_1^\top \Sigma e_1 = e_1^\top UU^\top e_1 = v^Tv,
\]
with $v:=U^\top e_1$. This is equivalent to solving the linear system 
$Lv=e_1$, which can be done efficiently using forward substitution.

In \cref{app:prior_samples} we show a few random samples from the prior
described in this section.

\section{Experimental setup}\label{sec:mcmc}

As discussed earlier, we use MCMC to sample from the posterior distribution given
by the computer model, the observed muon counts, and the prior. Here we detail
the specific algorithms used and the software involved in the implementation.

\subsection{Markov chain Monte Carlo sampling algorithms}

Given the nature of the model at hand, we expect to encounter challenges in the sampling 
process due to high correlation in the posterior distribution. These difficulties
stem from the fact that the height of one layer cannot be changed without altering
at least one other layer, so that the total height of the layer model remains the
same. Therefore, samplers that ignore these induced correlations will take a 
prohibitively long time to converge. We thus require advanced MCMC samplers that 
take into account the posterior geometry when proposing moves. 
In order to have access to such advanced MCMC samplers, we implement the model 
in NumPyro \citep{phan2019composable}, a Python-based probabilistic 
programming language (PPL; see e.g. \citealp{rainforth2017automating} and 
\citealp{vandemeent2021introduction}). PPLs allow researchers to write large 
classes of probabilistic models in a common language, and provide implementations
of state of the art MCMC samplers that can run off-the-shelf on any such model.
Compared to other PPLs, NumPyro has the additional benefit of being written in JAX 
\citep{jax2018github}, a Python framework with Numpy-like syntax that allows for 
automatic differentiation, vectorization, and just-in-time compilation that produces
efficient code for both CPUs and GPUs. Thus, NumPyro allows us to take advantage
of the recent increase in availability of compute-oriented GPUs with no additional
coding effort.

Automatic differentiation enables the use of advanced
gradient-based samplers that produce high-quality posterior samples with 
manageable cost, like Hamiltonian (or Hybrid) Monte Carlo (HMC, \citealp{Duane87}; 
see also \citealp{Neal11,betancourt2018conceptual}). Let us call \emph{latent vector} 
the result of reshaping (as in \cref{fig:array_indexing}) and stacking all the 
unknown variables in the model into a single one-dimensional array. 
Roughly speaking, HMC identifies
the latent vector with the position of a fictitious particle, which it associates
with a momentum variable. By carefully designing an energy function in phase space, 
the trajectories given by exactly integrating Hamilton's equations interspersed with
random momentum perturbations produce (correlated) samples from the target 
distribution. 

In practice, however, numerical integration is required to solve 
for HMC trajectories, which in turn forces the introduction of an accept-reject 
step in order to compensate for the loss of accuracy. The integrator has two
parameters: a positive step size, and the number of steps to take at each iteration.
Finding adequate values for these parameters is a hard problem. In our experiments
we use the no-u-turn sampler (NUTS) introduced in \citet{HoffmanGelman14}, an 
adaptive version of HMC. At each iteration, NUTS finds an adequate number of 
integration steps to take so as to avoid backtracking by stopping after 
encountering a so called \emph{u-turns}. Backtracking arises due to the fact that,
under mild conditions, the constant-energy orbits of the Hamiltonian system 
are periodic. Therefore, the process returns to its initial state in finite time. 
By identifying the point at which the distance to the origin starts to 
decrease---i.e., a u-turn---NUTS stops integrating past that point to avoid
wasting computational resources, resulting in an overall more efficient sampler.
Additionally, NUTS defines a robust adaptation procedure to tune the step 
size parameter, targeting an average Metropolis acceptance probability of
$80\%$.

\subsection{Multiple independent chains via vectorization}

We take advantage of automatic vectorization to run several chains
simultaneously on the same GPU. Running independent chains with random initializations
is an easy way to improve the exploration of the posterior distribution.
To avoid performance degradation of the vectorized sampler caused by 
MCMC steps with widely varying random duration \citep[see e.g.][]{dance2025efficiently}%
---which are common when employing adaptive samplers such as NUTS---we lower the 
maximum trajectory length that NUTS is allowed to evaluate to the lowest 
possible value that does not result in sample quality degradation. We find that 
a maximum trajectory of $256$ steps strikes a good balance. In this regime, 
almost all NUTS iterations reach the maximum trajectory length---and therefore take 
roughly the same time to complete---resulting in little to no degradation of the 
vectorized sampler.

Notwithstanding the above, naively running independent chains with arbitrary 
initialization can give the impression of mixing when in fact each chain ends up
stuck in a subset of the state space dictated entirely by the (arbitrary) 
initial state \citep{cowles1996markov}. 
In order to detect the presence and severity of this effect, we employ the 
\emph{nested}-$\hat{R}$ diagnostic introduced by \citet{margossian2024nested}.
This statistic is a generalization and improvement of the classical $\hat{R}$
or Gelman--Rubin diagnostic \citep{gelman1992inference}. It relies on a novel
experimental design for running multiple MCMC processes, where a total of
$\nchains:=\nsuper\cdot\nwithin$ chains are grouped into $\nsuper$ 
\emph{super-chains}, each with $\nwithin$ components. Every super-chain is 
assigned a randomly sampled initial state, and all its $\nwithin$ component 
chains start from this state. In an ideal setting, every chain is able to 
explore the full posterior distribution, regardless of their initial state. 
Therefore, samples from different super-chains should be indistinguishable.
In the opposite case, chains are stuck in a neighborhood close to their 
starting point, and thus samples from different super-chains look nothing alike.

For the experiments in the next section, we use $\nsuper=32$ super-chains
containing $\nwithin=8$ chains each. Depending on the number of available 
GPUs and their memory capacity, chains can all be run vectorized on the same 
device, or vectorized only within a super-chain and then either distributed 
across non-communicating GPUs, or sequentially ran on the same device. In our
experiments we run all $\nchains$ on the same GPU.

Furthermore, following the suggestions in \citet{margossian2024nested} and the 
many short chains principle for GPU vectorized MCMC \citep{sountsov2026running},
we discard as warm-up (or burn-in) all but the last step of each chain, leaving a 
total of $\nchains=256$ independent samples approximately distributed according to 
the posterior. This has negligible impact on the sample quality (due to the 
presence of high auto-correlation) but it allows us to increase the number
of chains that run simultaneously (as we only require keeping one copy of the 
state of each chain in GPU memory as opposed to the whole trajectory). It also
speeds up sampling since storing every chain step necessarily requires copy
operations.

In our experiments, each chain first runs $2^{13}$ adaptation steps---during 
which the
step size of the integrator is automatically tuned, along with a diagonal 
preconditioner---followed by an equal number of sampling steps, of which only the
last one is retained (see \cref{app:long_mcmc} for a justification of these
numbers).
Finally, we note that if an application required increased precision and thus 
additional independent samples, the ideal way to produce them is by increasing
$\nsuper$ while keeping the rest of the configurations above fixed.

\section{Simulation experiment}\label{sec:experiment}

As we described in \cref{sec:model}, the block-caving example visualized in 
\cref{fig:true_layers} corresponds to a simulated cave model, which means we 
have access to the true heights $H^\text{true}$. We can use this to compute the 
expected counts $\lambda(\hat R(H^\text{true}))$ under this geometry for the
sensors shown in \cref{fig:true_layers}. If we round these expected counts
to the nearest integer, we can treat these values as observed counts, which then
allows us to empirically assess if our proposed framework is able to produce
samples that are compatible with $H^\text{true}$. In this experiment, each of the
$25$ sensors contains $\npps=16^2=256$ pixels, which amounts to a total of
$\npixels=6400$ pixels. We simulate an exposure time of $30$ days, keeping the
geometry fixed.

\begin{myfig}
\centering
\begin{subfigure}[b]{0.5\myfigwidth}
\includegraphics[width=\textwidth]{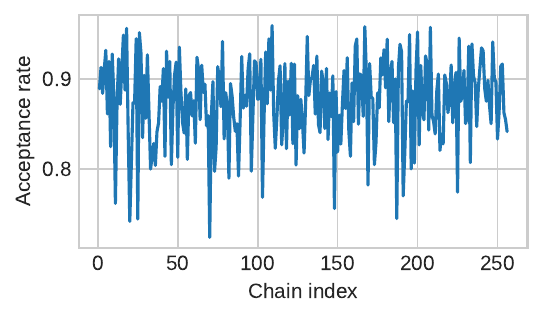}
\vspace{-0.6cm}
\caption{Post-warmup avg. acceptance probability.}
\label{fig:traces_acc_prob_poisson}
\end{subfigure}%
\begin{subfigure}[b]{0.5\myfigwidth}
\includegraphics[width=\textwidth]{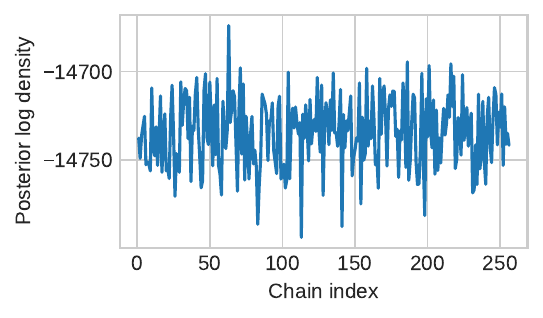}
\vspace{-0.6cm}
\caption{Posterior log density at last step.}
\label{fig:traces_log_density_poisson}
\end{subfigure}\\[0.2cm]
\begin{subfigure}[b]{0.5\myfigwidth}
\includegraphics[width=\textwidth]{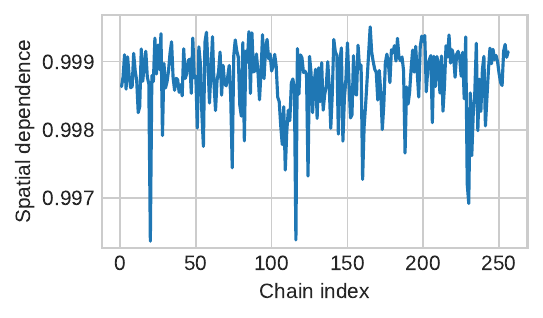}
\vspace{-0.6cm}
\caption{Layer 1 spatial dependence $r_1$ at last step.}
\label{fig:traces_rho_1_poisson}
\end{subfigure}%
\begin{subfigure}[b]{0.5\myfigwidth}
\includegraphics[width=\textwidth]{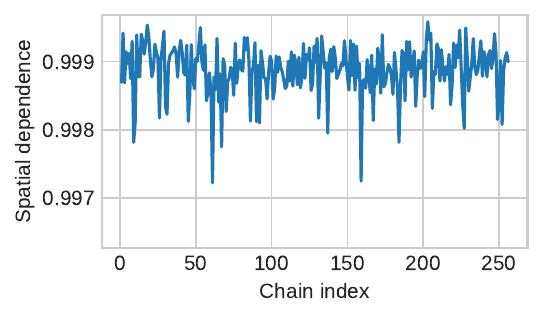}
\vspace{-0.6cm}
\caption{Layer 2 spatial dependence $r_2$ at last step.}
\label{fig:traces_rho_2_poisson}
\end{subfigure}
\caption{Visualization of relevant statistics for the $256$ chains.
Samples from the same super-chain are plotted contiguously.}
\label{fig:traces_poisson}
\end{myfig}

After running the ensemble of independent chains as described in
\cref{sec:mcmc} and keeping only their last steps, we find that the worst
(i.e., highest) nested-$\hat{R}$ across all latent variables is $1.12$---which
is close to the ideal of $1$---implying that no obvious convergence issue is
present. We visually corroborate this in \cref{fig:traces_poisson}, plotting 
several sample statistics for the $256$ chains that comprise the full MCMC run,
such that the ones belonging to the same super-chain are plotted contiguously.
Consequently, if these groups were visually apparent in the plots, we would
suspect there are issues with mixing. In contrast, the four panels in 
\cref{fig:traces_poisson} show values that seem drawn from the 
same distribution. Moreover, \cref{fig:traces_acc_prob_poisson} indicates that
all of the chains achieve average post-warmup acceptance probabilities that are
close to the $80\%$ target for NUTS. We also notice in
\cref{fig:traces_rho_1_poisson,fig:traces_rho_2_poisson} that all the chains show 
layer smoothness values $(r_1,r_2)$ close to $1$, indicating that the sampled
layer models evidence spatial dependence. This validates the modeling choice
of replacing the independent columns prior with structured layers, as described
in \cref{sec:CAR_prior}.

\begin{myfig}
\centering
\includegraphics[width=0.25\myfigwidth,trim={165pt 0pt 15pt 15pt},clip]
{true_layers_0}%
\includegraphics[width=0.25\myfigwidth,trim={165pt 0pt 15pt 15pt},clip]
{true_layers_3}%
\includegraphics[width=0.25\myfigwidth,trim={120pt 0pt 60pt 15pt},clip]
{true_layers_2}%
\includegraphics[width=0.25\myfigwidth,trim={15pt 0pt 165pt 15pt},clip]
{true_layers_1}\\
\includegraphics[width=0.25\myfigwidth,trim={165pt 0pt 15pt 0pt},clip]
{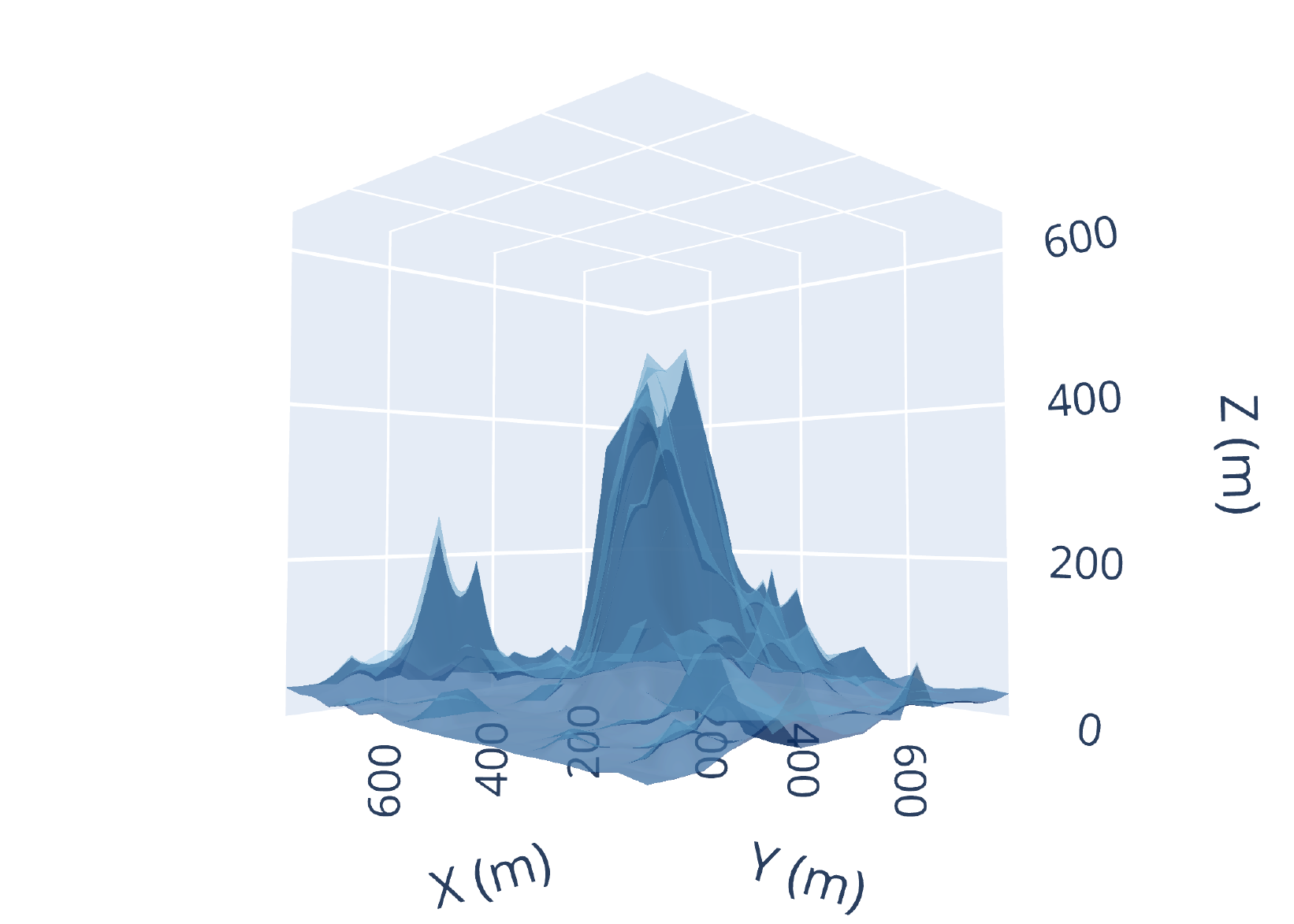}%
\includegraphics[width=0.25\myfigwidth,trim={165pt 0pt 15pt 0pt},clip]
{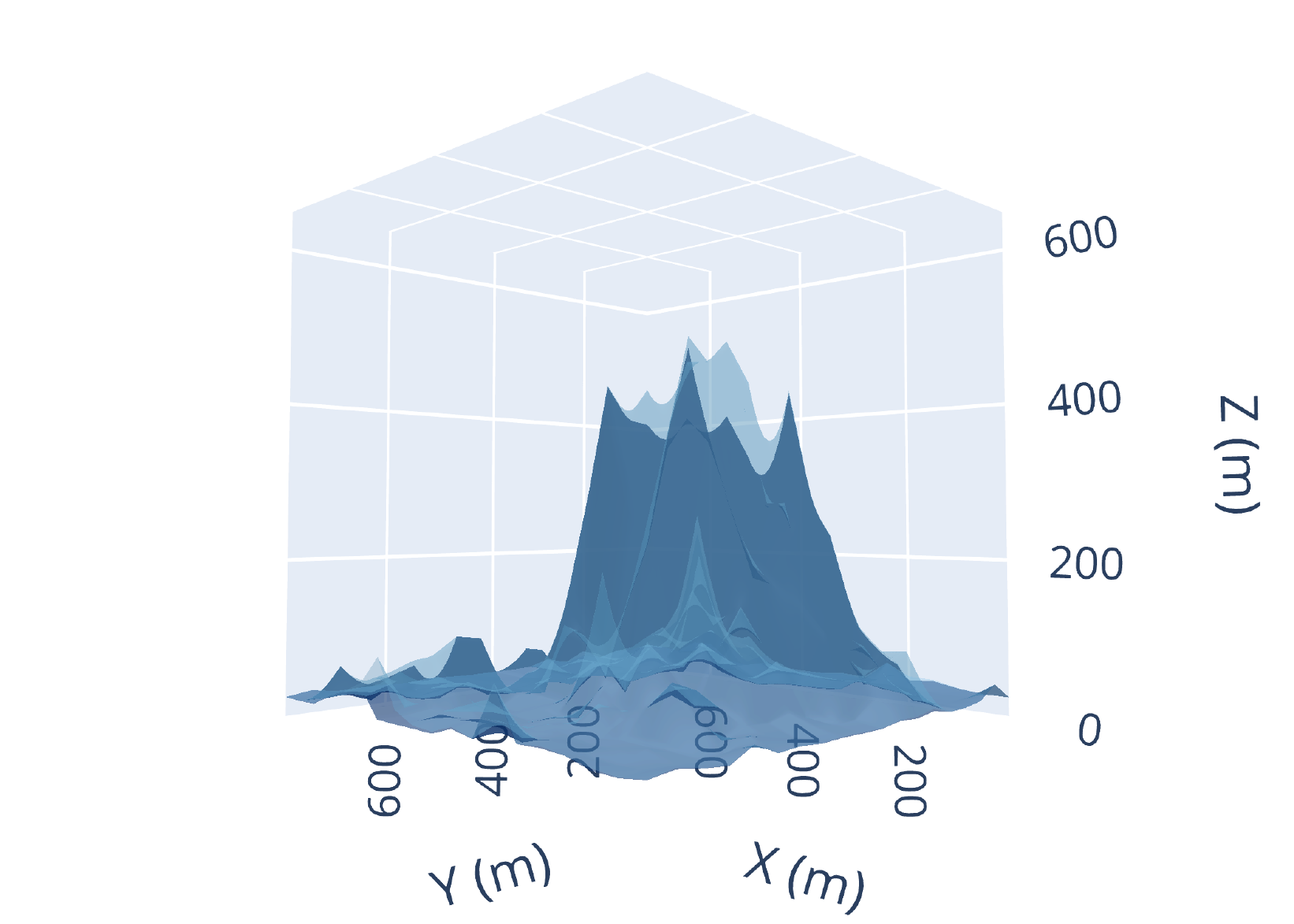}%
\includegraphics[width=0.25\myfigwidth,trim={120pt 0pt 60pt 0pt},clip]
{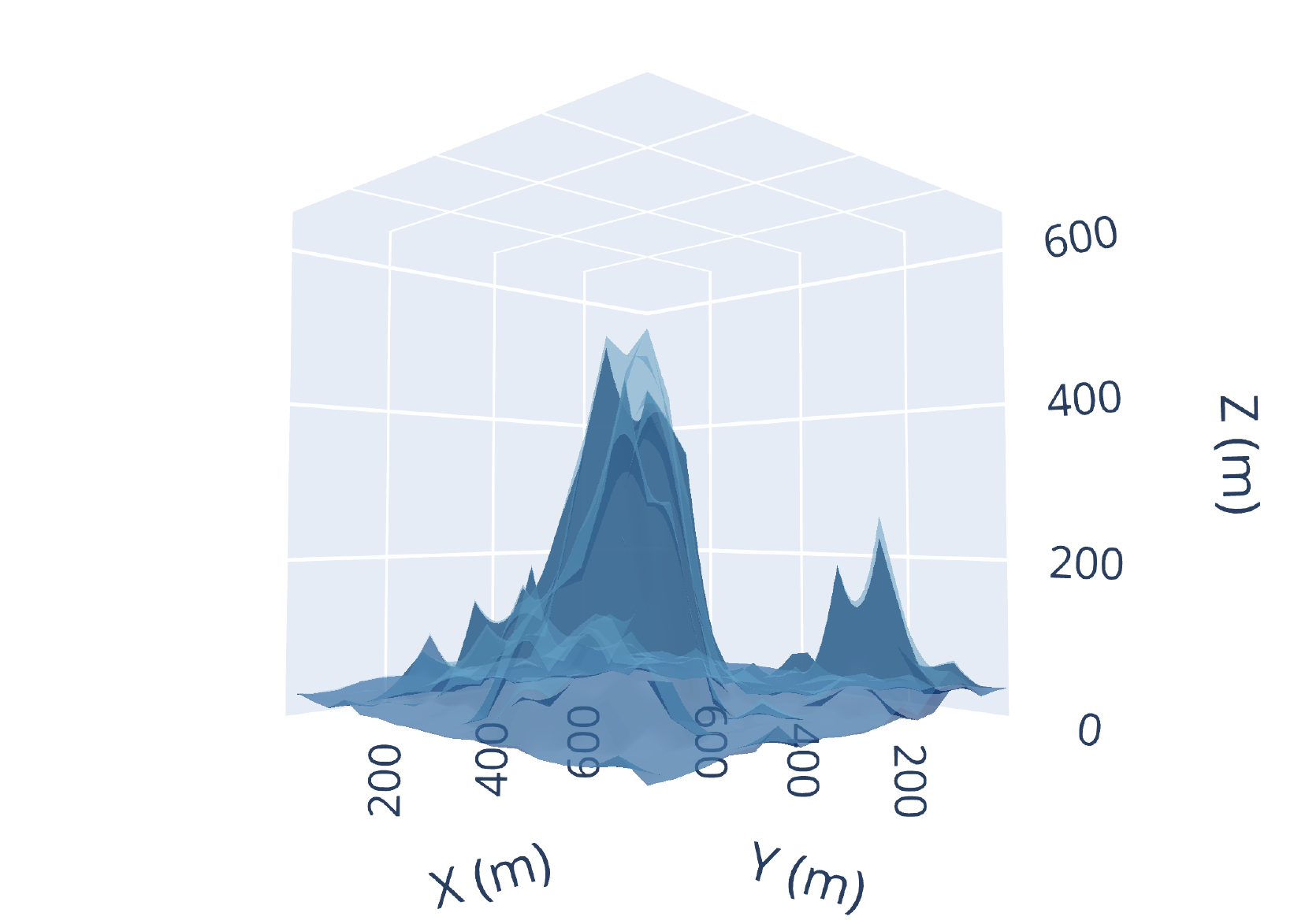}%
\includegraphics[width=0.25\myfigwidth,trim={15pt 0pt 165pt 0pt},clip]
{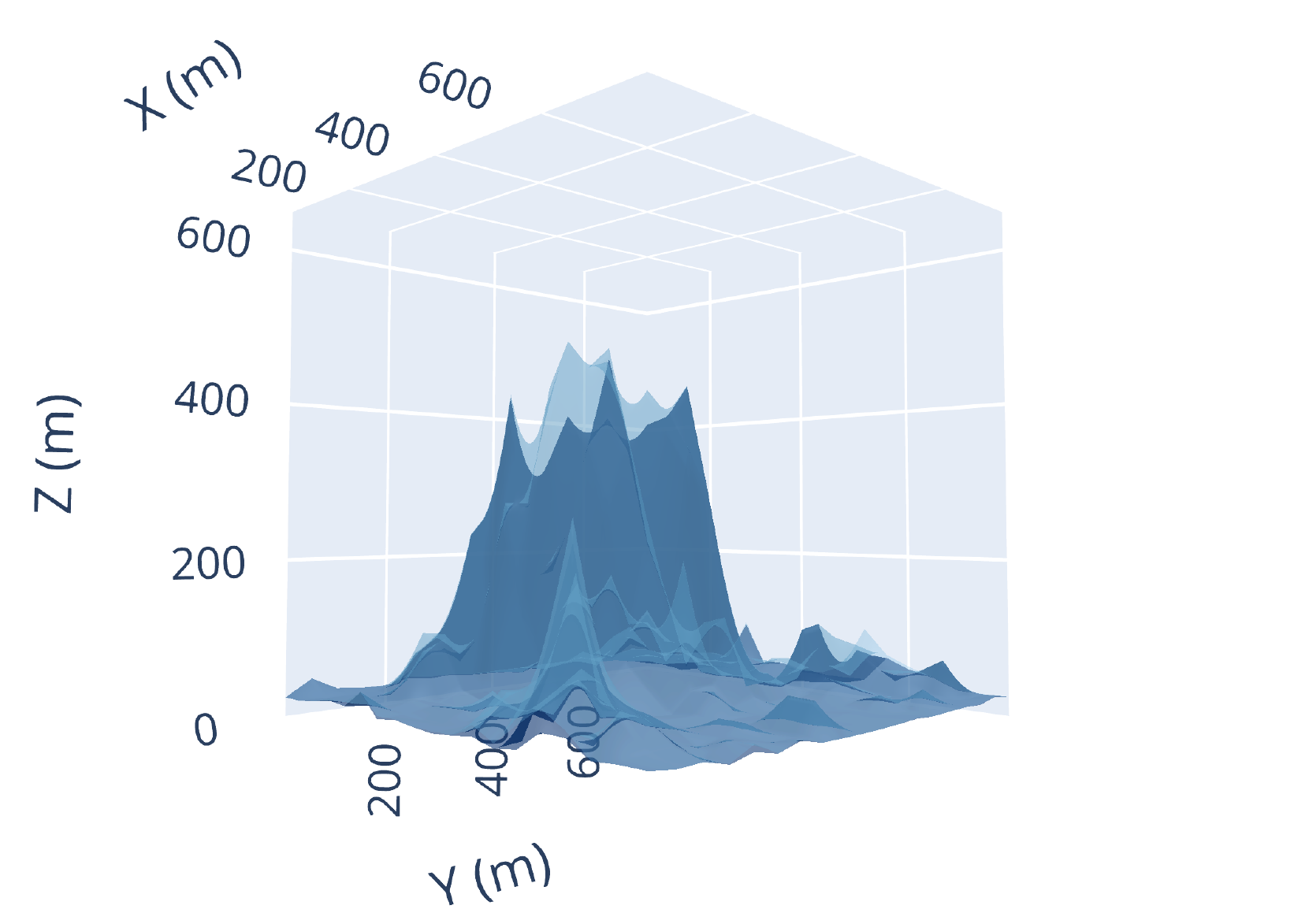}\\
\includegraphics[width=0.25\myfigwidth,trim={165pt 10pt 15pt 0pt},clip]
{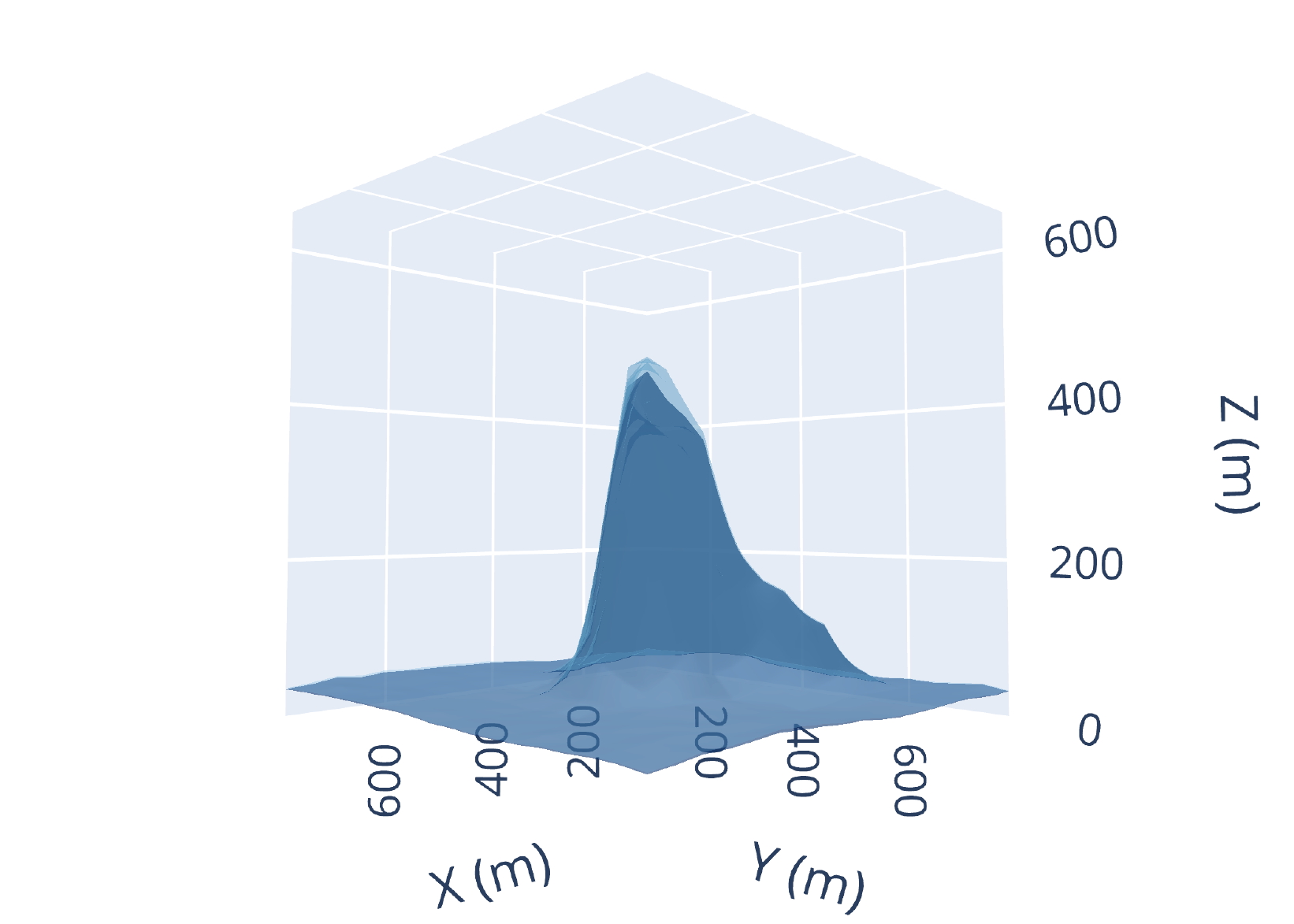}%
\includegraphics[width=0.25\myfigwidth,trim={165pt 10pt 15pt 0pt},clip]
{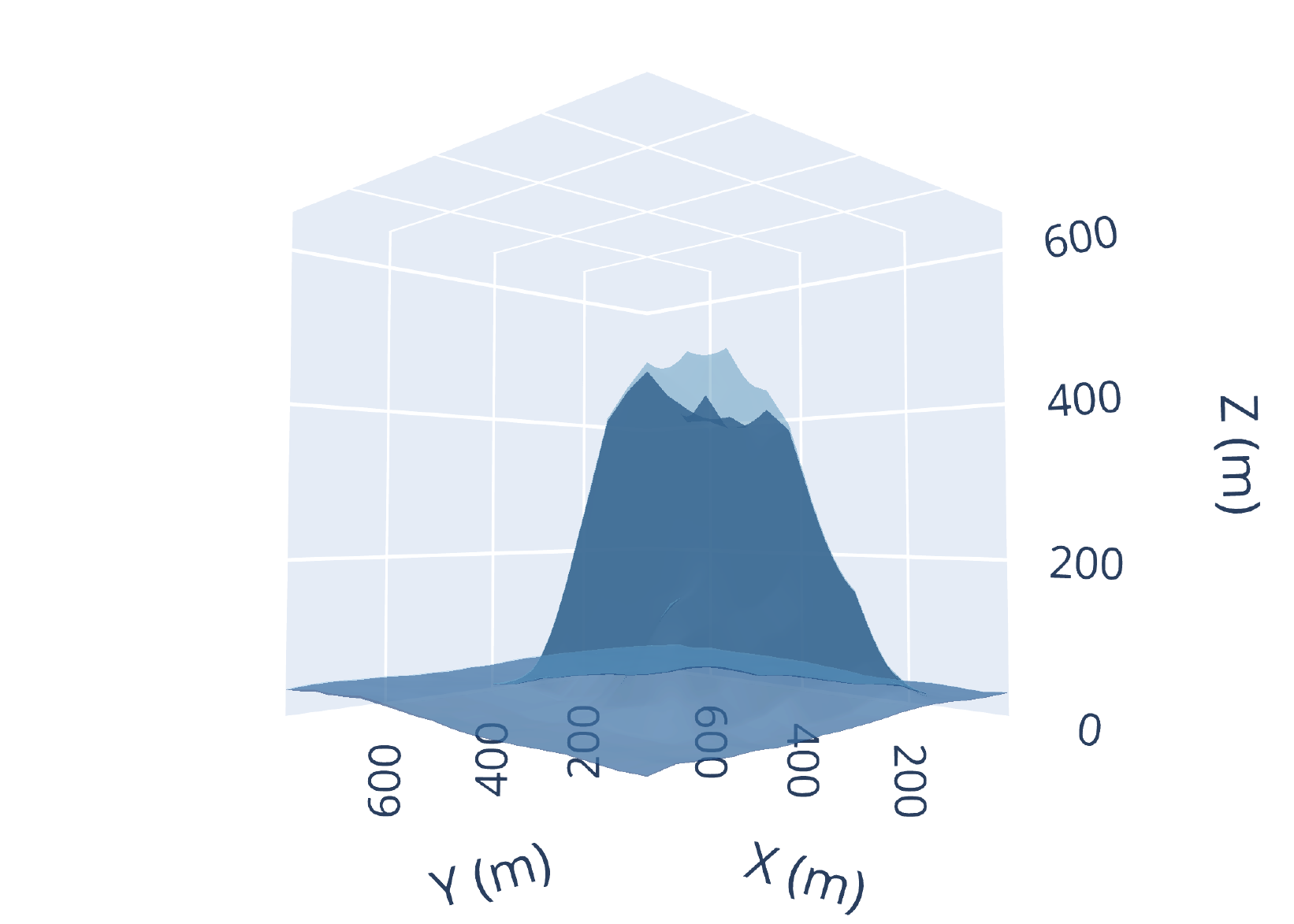}%
\includegraphics[width=0.25\myfigwidth,trim={120pt 10pt 60pt 0pt},clip]
{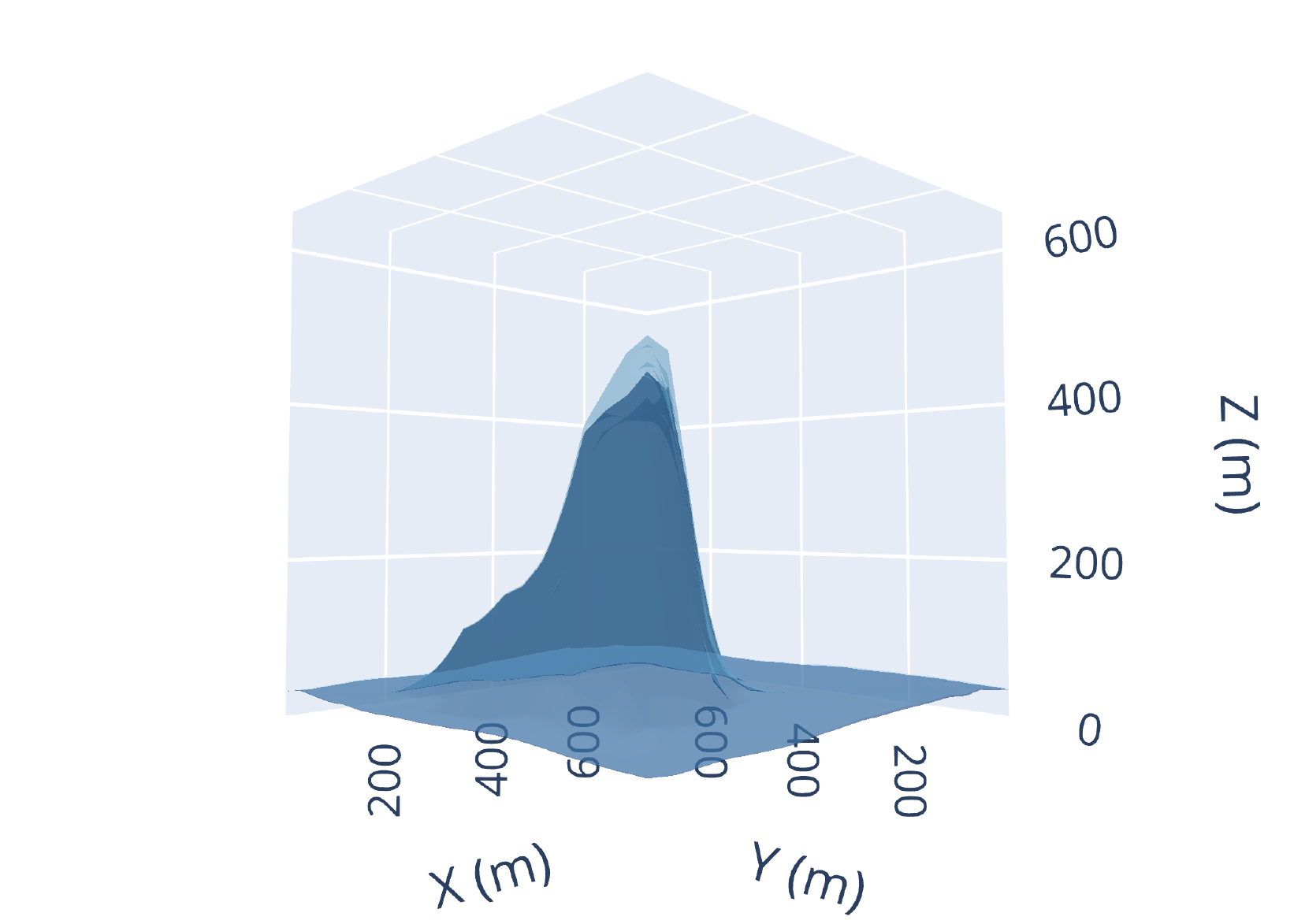}%
\includegraphics[width=0.25\myfigwidth,trim={15pt 10pt 165pt 0pt},clip]
{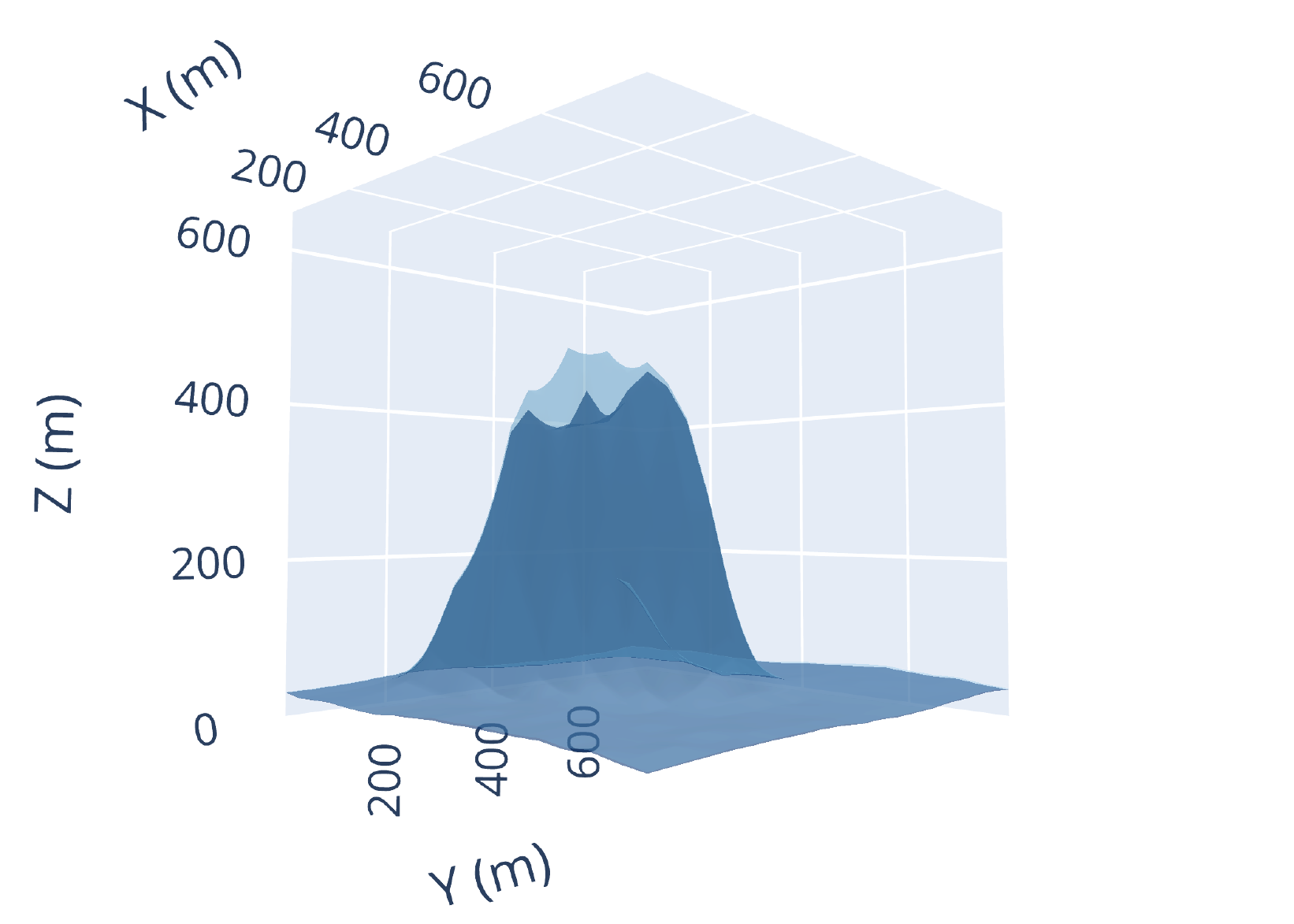}
\caption{Comparing two point estimators for the layer heights against the truth.
Successive plots from left to right correspond to $90^\circ$ clockwise rotations
about the $z$ axis.
\textbf{Top}: true layers (same as \cref{fig:true_layers}). 
\textbf{Middle}: sample with the highest posterior density (MAP). 
\textbf{Bottom}: posterior mean of the layer heights.}
\label{fig:point_estimates_poisson}
\end{myfig}

\cref{fig:point_estimates_poisson} shows a visualization of two
point estimates of the cave geometry: the sample estimate of the maximum 
\emph{a posteriori} (MAP)---i.e., the sample with the highest posterior density---%
and the posterior mean of the layer heights (or, equivalently, the thicknesses, as
these are related by a linear transformation). The MAP looks like a noisy version
of the true geometry, such that its overall shape and features are recovered. 
The spikiness of the MAP evidences the uncertainty in the exact position of the 
layers. We expect this uncertainty to be higher in regions with no sensors, as
exemplified by the fictitious secondary cave shown in the first two plots of the
second row of \cref{fig:point_estimates_poisson}.
On the other hand, the posterior mean averages out this type of noise, 
resulting in a geometry that is closer to the truth than any specific sample.

\begin{myfig}
\centering
\begin{subfigure}[b]{\myfigwidth}
\includegraphics[width=\textwidth,trim={5pt 0pt 10pt 5pt},clip]{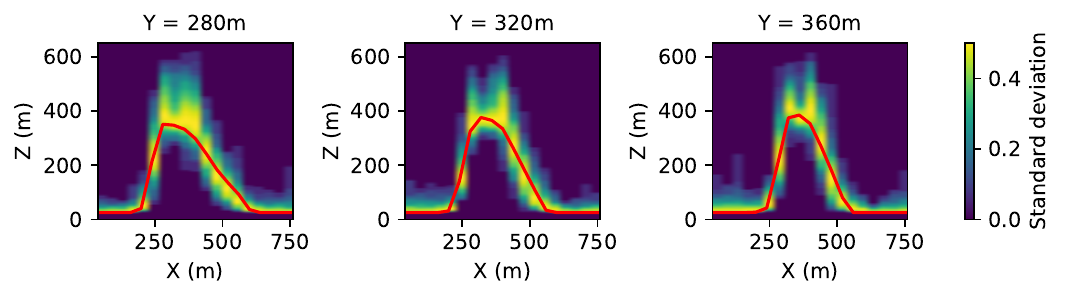}
\vspace{-0.7cm}
\caption{Standard deviation of the $B_{i,j,k}^{(1)}(H)$ indicators versus true 
muck-air interface}
\label{fig:tomography_std_in_muckpile_poisson}
\end{subfigure}\\[0.2cm]
\begin{subfigure}[b]{\myfigwidth}
\includegraphics[width=\textwidth,trim={5pt 0pt 10pt 0pt},clip]{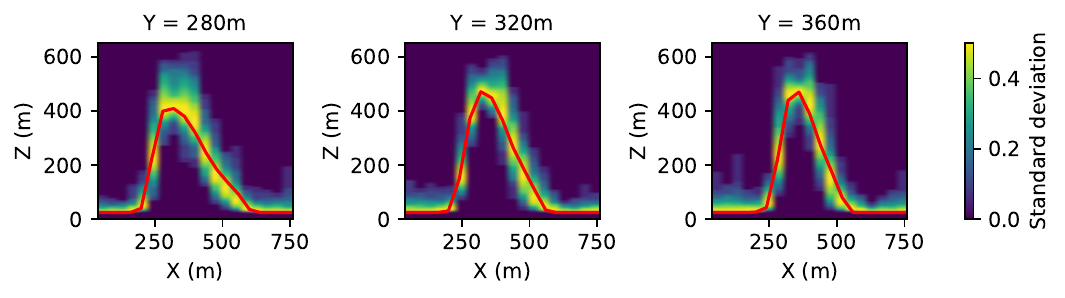}
\vspace{-0.7cm}
\caption{Standard deviation of the $B_{i,j,k}^{(2)}(H)$ indicators versus true 
air-rock interface.}
\label{fig:tomography_std_in_cave_poisson}
\end{subfigure}
\caption{Tomographic visualization across three selected slices along the $y$
axis close to the peak of the cave, showing the standard deviation across MCMC 
samples of below-interface indicators $B_{i,j,k}^{(\ell)}(H)$ 
(\cref{eq:def_below_interface_indicator}). Red lines correspond to true interfaces.
}
\label{fig:tomography_std_poisson}
\end{myfig}

Still, the volume of air seems under estimated in both summaries shown in
\cref{fig:point_estimates_poisson}. In contrast to such point estimates, adequate 
risk assessment should also provide uncertainty measures. To this
end, \cref{fig:tomography_std_poisson} shows a tomographic visualization of the 
uncertainty in the predicted layer locations, across three selected slices along 
the $y$ axis of the domain $\mathcal{D}$ where the bulk of the cave is located.
Specifically, the heat maps show the 
standard deviation across MCMC samples of the indicators $B_{i,j,k}^{(\ell)}(H)$
for a voxel $(i,j,k)$ lying below the interface between layers $\ell$ and
$\ell+1$ (\cref{eq:def_below_interface_indicator}). As the standard deviation of
any binary random variable, the metric is contained in $[0,0.5]$, where $0$ 
means that the indicator is constant across samples, while $0.5$ corresponds to
the indicator taking the values $0$ and $1$ the same number of times. The latter 
case suggests the likely locations of the layer interfaces. Similar 
visualizations have been used in the literature to provide uncertainty 
quantification for interface locations
\citep[see e.g.][]{wang2023unraveling}. Furthermore, we overlay the images with 
the true layer boundaries in order to check if the MCMC samples are consistent
with the ground truth.

\cref{fig:tomography_std_in_muckpile_poisson} focuses on the muck-air
interface, where we see that the region of maximal uncertainty contains
the ground truth interface, while the indicators quickly stabilize away from it.
However, in the region closer to the peak (around $x=320$), the plots show
substantial probability that the interface is much higher.
On the other hand, \cref{fig:tomography_std_in_cave_poisson} shows 
that the maximal uncertainty region for the air-rock interface is smaller and
more tightly situated around the true boundary. The net effect is consistent
with the fact that the point estimates in \cref{fig:point_estimates_poisson}
slightly underestimate the size of the air-gap, since the muck-air interface is
on average closer to the air-rock interface than in the ground truth model.
Notwithstanding the above, the fact that the true boundaries 
are seen within the high uncertainty regions in \cref{fig:tomography_std_poisson}
suggests that our methodology is capable of providing non-trivial uncertainty
quantification for the existence of air-gaps.

\section{Conclusions}

In this work we described a Bayesian model for muon tomography, along with a procedure
for performing inference via MCMC. We described how our
implementation leveraged recently developed computational frameworks that simplify 
the description of the model, offer off-the-shelf modern MCMC samplers, and 
allow seamless execution of the inference process on GPUs. We study the properties
of our method in a simulated block caving dataset, where the true geometry is known.
These experiments suggest that our proposed MCMC scheme is able to adequately explore
the space of geometries defined by our parametrization of the problem. In particular, 
these simulations suggest that the methodology is able to capture the risk of air
gap at the top of the cave.

The results suggest several potential avenues of future research. For instance,
our probabilistic methods can be readily expanded from a static, single-time
modeling approach to the continuous monitoring setting. Indeed, using the prior
designed in this work as an initial distribution, together with a probabilistic 
time-stepping mechanism for the expected evolution of the cave interfaces, we
would obtain a sequential Monte Carlo \citep[see e.g.][]{chopin2020introduction}
algorithm for uncertainty quantification that could exploit the time dimension
of the muon count process---and thus account for the evolution of the cave
geometry.

On the computational
side, we note that the bottleneck of our model lays in the matrix manipulations required
to simulate layers of correlated heights. In the future, it would be interesting to 
consider alternative specifications of the underlying Gaussian random fields and/or
of simulation techniques which might speed-up the evaluation of the posterior density.
In terms of the modeling approach, it is important to note that layer models cannot
express the totality of cave geometries that can arise (e.g., hourglass-shaped).
In the future, we would like to explore the use of level sets to describe more complex
geometries \citep[see e.g.][]{xie2011uncertainty,iglesias2016bayesian}.
 
\section*{Acknowledgments}

The authors wish to thank Tony Diering for kindly providing us with the simulated
block caving example we use throughout this work.
Derek Bingham acknowledges the support of the Natural Sciences and Engineering 
Research Council of Canada.
Donald Estep's work has been partially supported by the Canada Research Chairs Program,
the Natural Sciences and Engineering Research Council of Canada, Canada's Digital 
Technology Supercluster, the Dynamics Research Corporation, the Idaho National 
Laboratory, the United States Department of Energy, the United States National 
Institutes of Health, the United States National Science Foundation, and Riverside 
Research.

\section*{Code availability}

A GitHub repository has been set up for the algorithms and Python code 
developed in this work:
\url{https://github.com/Estep-Bingham-Lab/layermodels-repr}.

\bibliography{main}

\appendix
\counterwithin{figure}{section} 
\section{Layer models sampled from the prior}\label{app:prior_samples}

\begin{myfig}
\centering
\includegraphics[width=0.25\myfigwidth,trim={165pt 0pt 15pt 15pt},clip]
{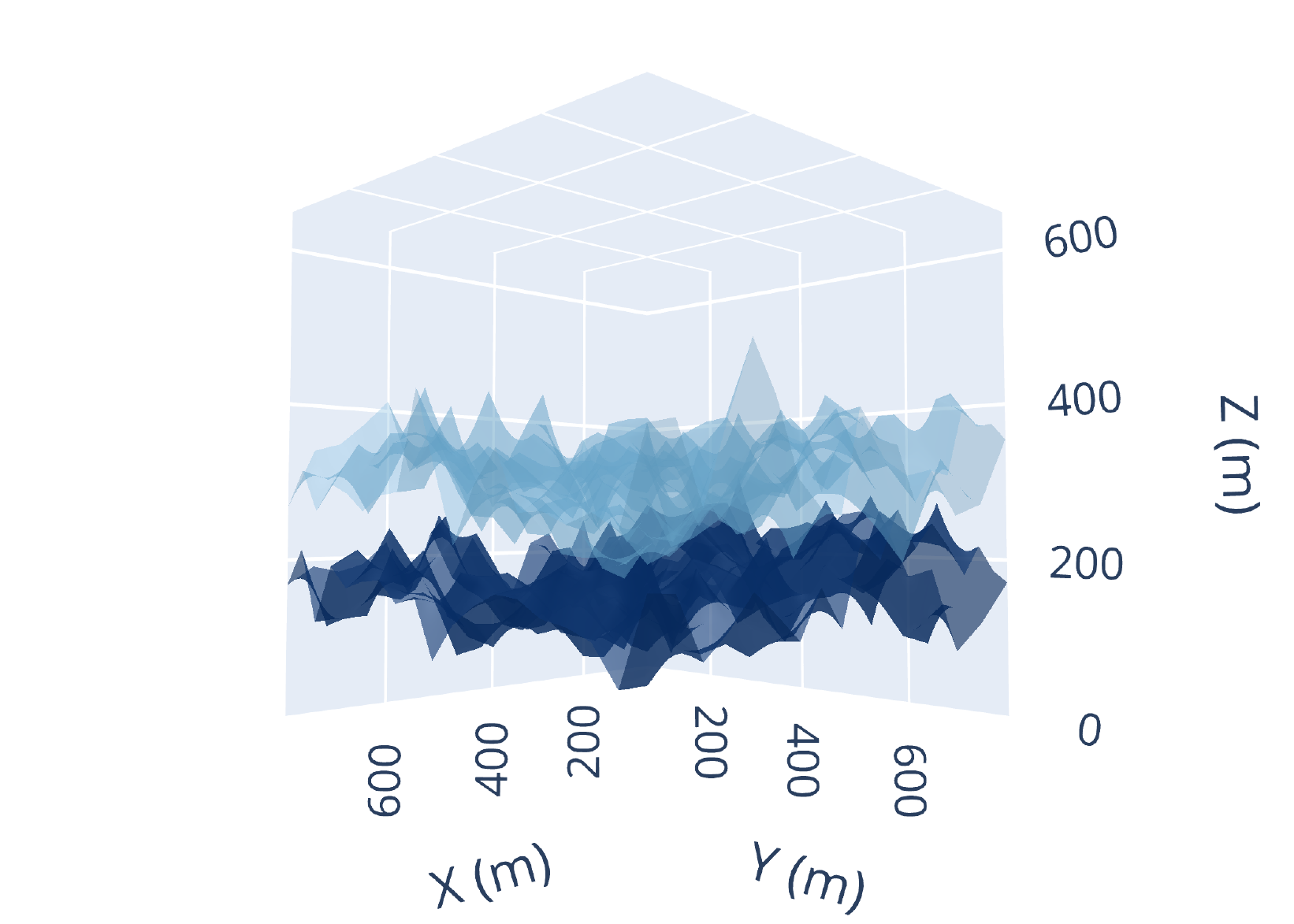}%
\includegraphics[width=0.25\myfigwidth,trim={165pt 0pt 15pt 15pt},clip]
{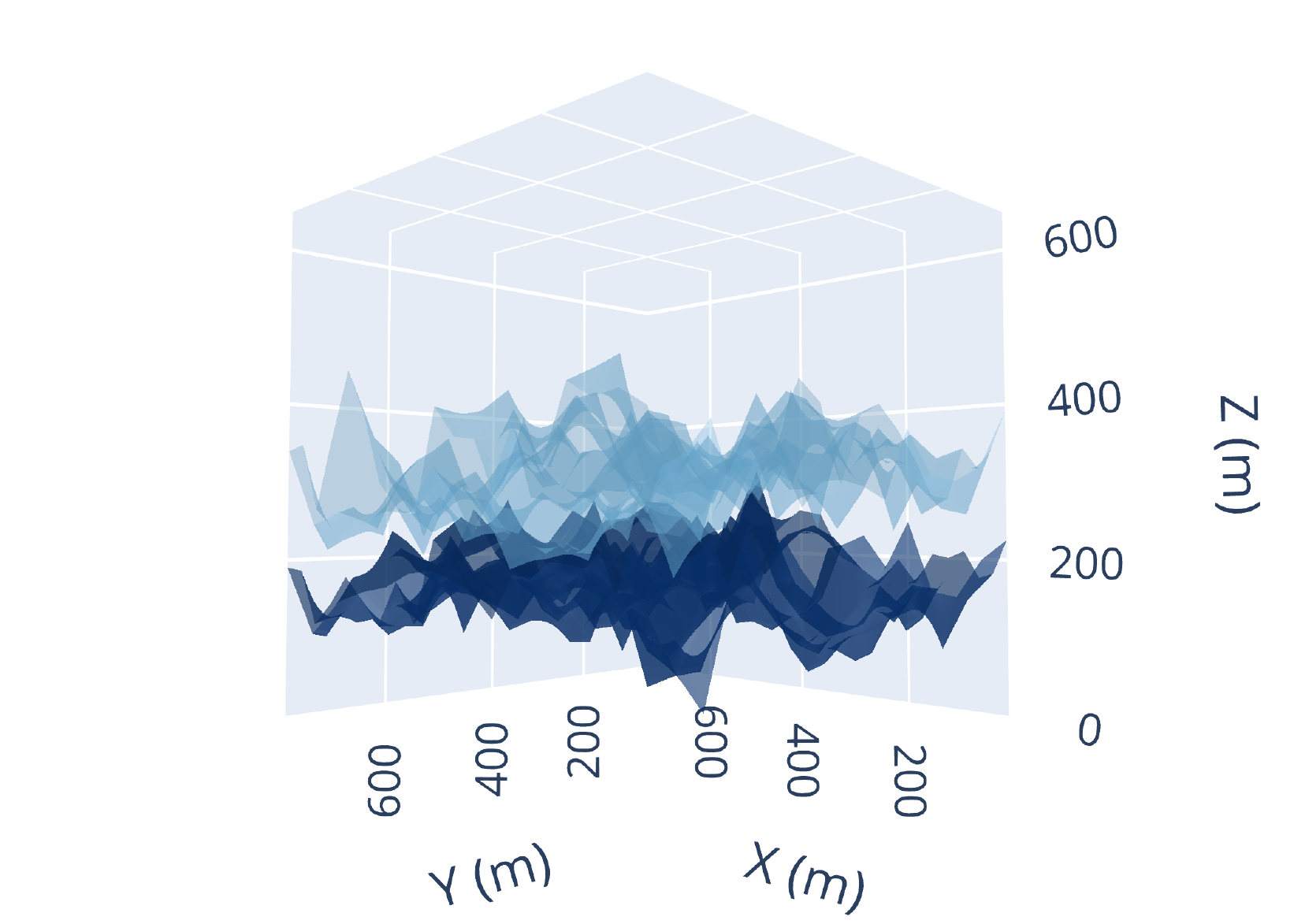}%
\includegraphics[width=0.25\myfigwidth,trim={120pt 0pt 60pt 15pt},clip]
{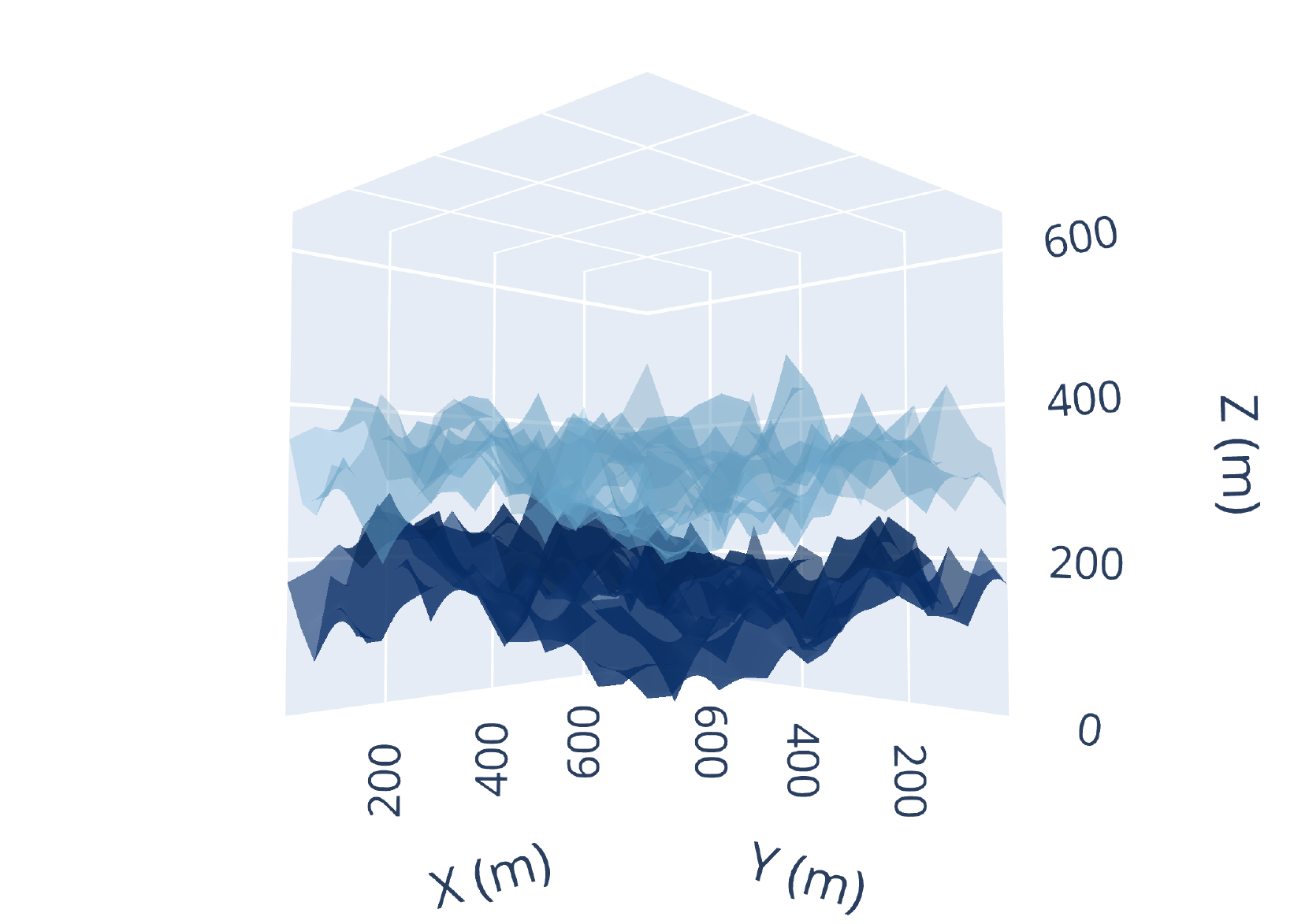}%
\includegraphics[width=0.25\myfigwidth,trim={15pt 0pt 165pt 15pt},clip]
{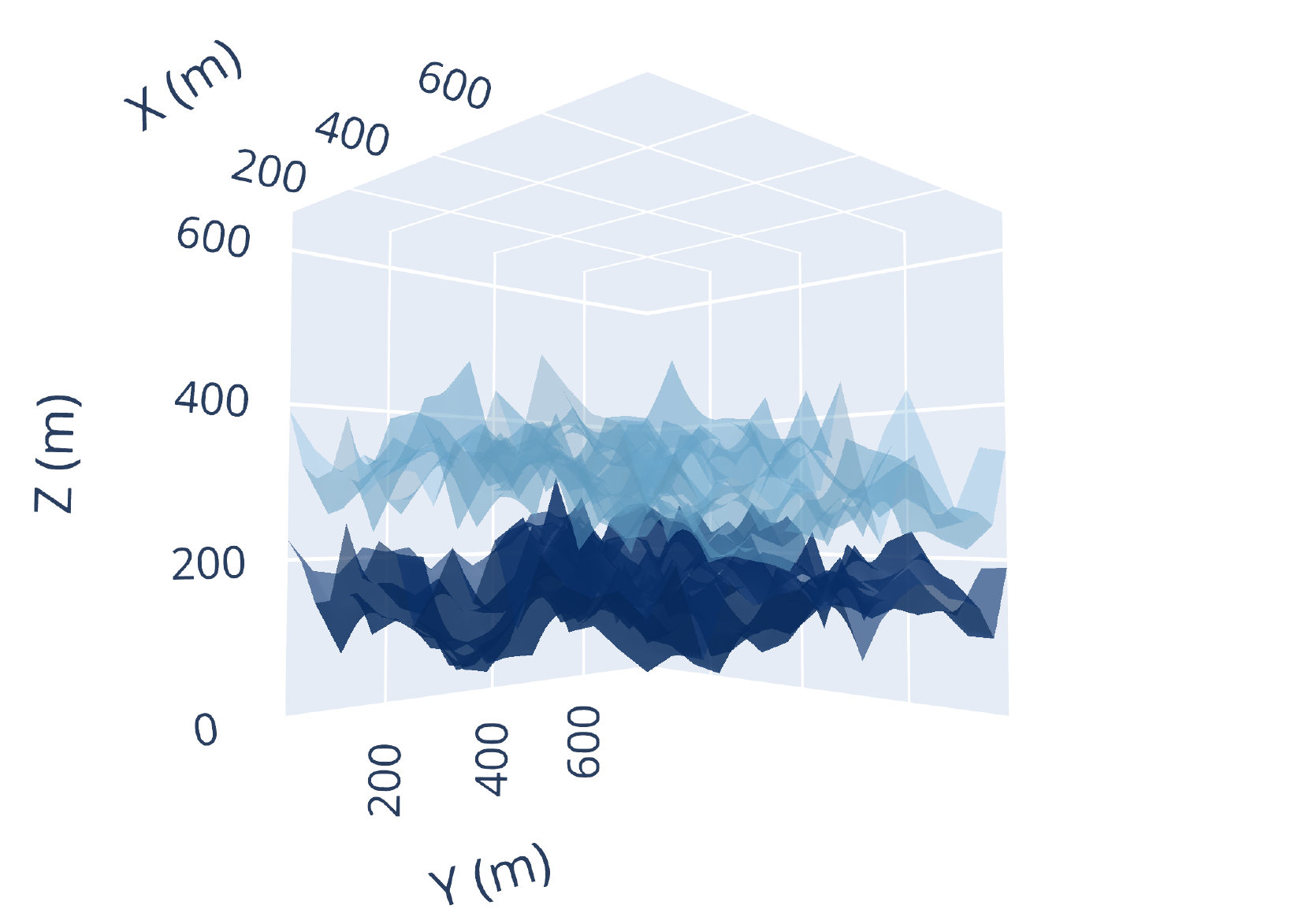}\\
\includegraphics[width=0.25\myfigwidth,trim={165pt 0pt 15pt 0pt},clip]
{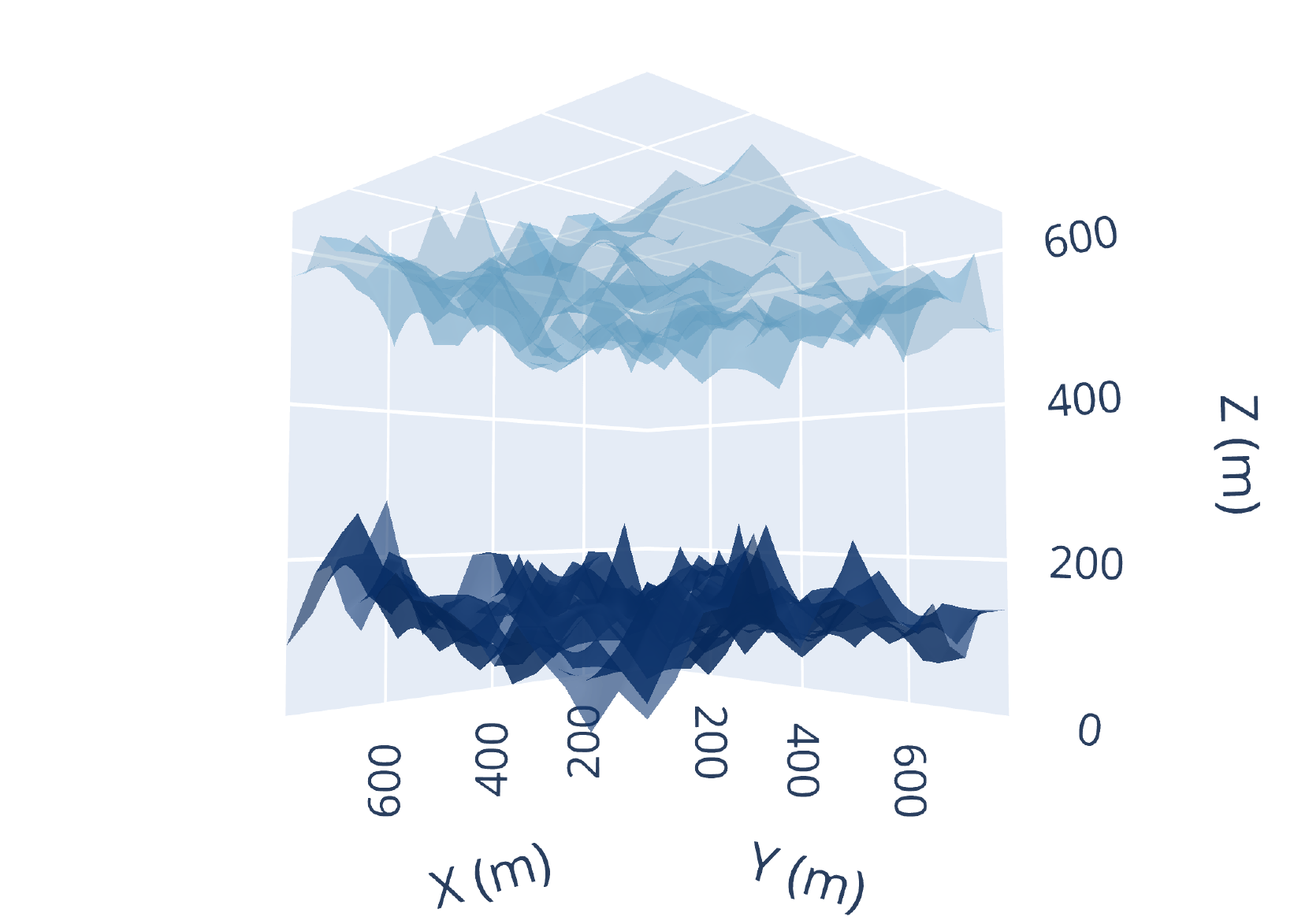}%
\includegraphics[width=0.25\myfigwidth,trim={165pt 0pt 15pt 0pt},clip]
{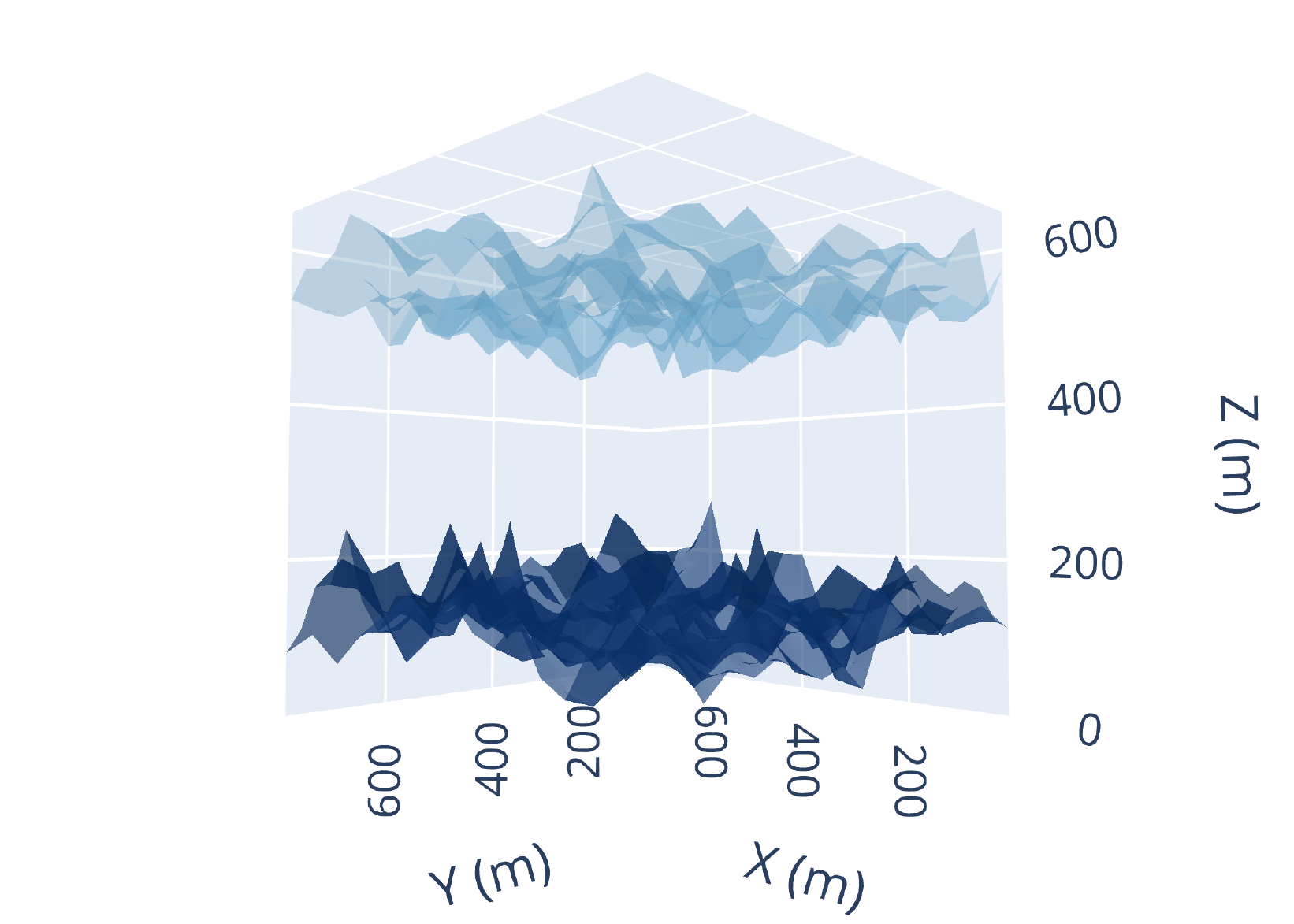}%
\includegraphics[width=0.25\myfigwidth,trim={120pt 0pt 60pt 0pt},clip]
{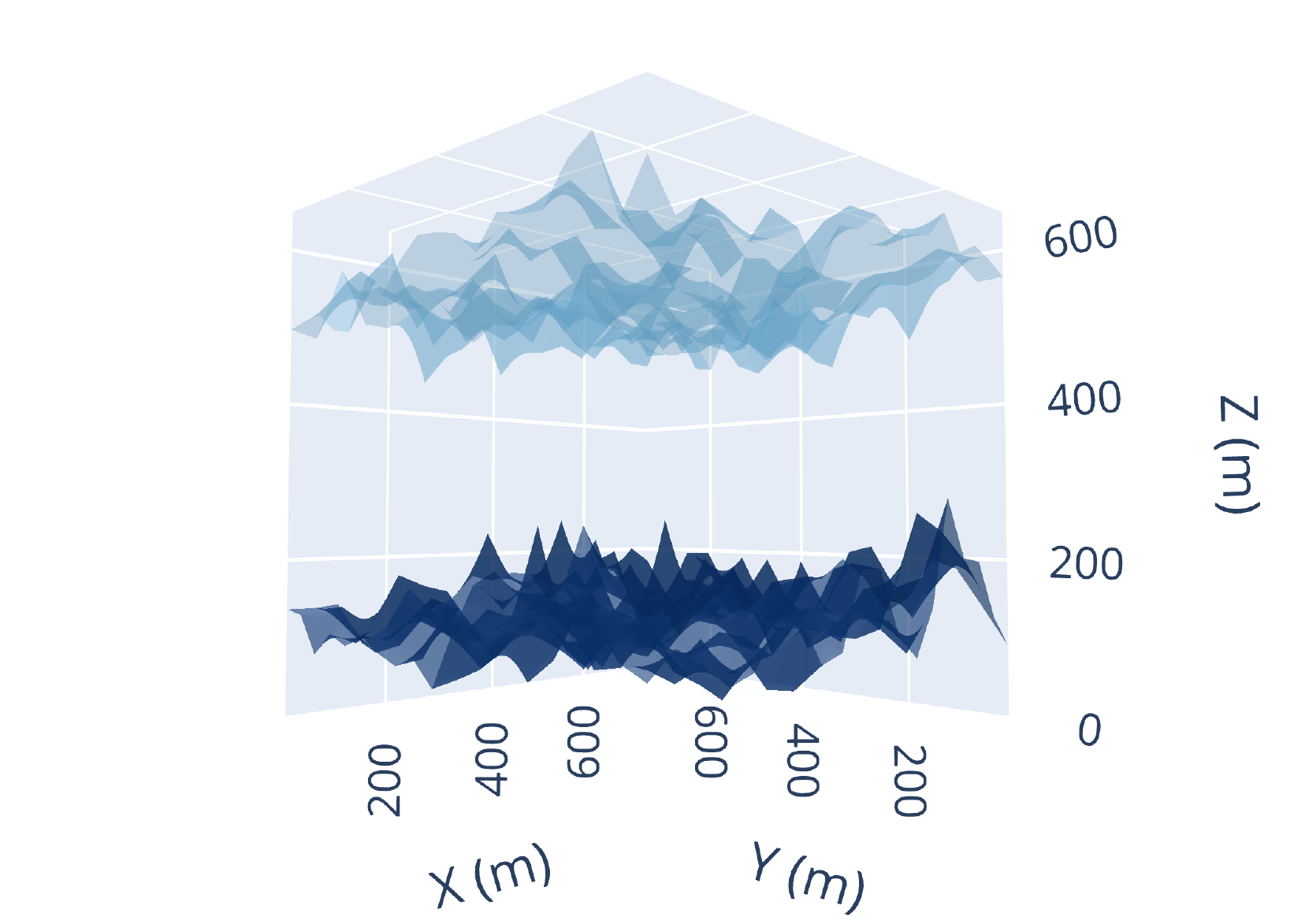}%
\includegraphics[width=0.25\myfigwidth,trim={15pt 0pt 165pt 0pt},clip]
{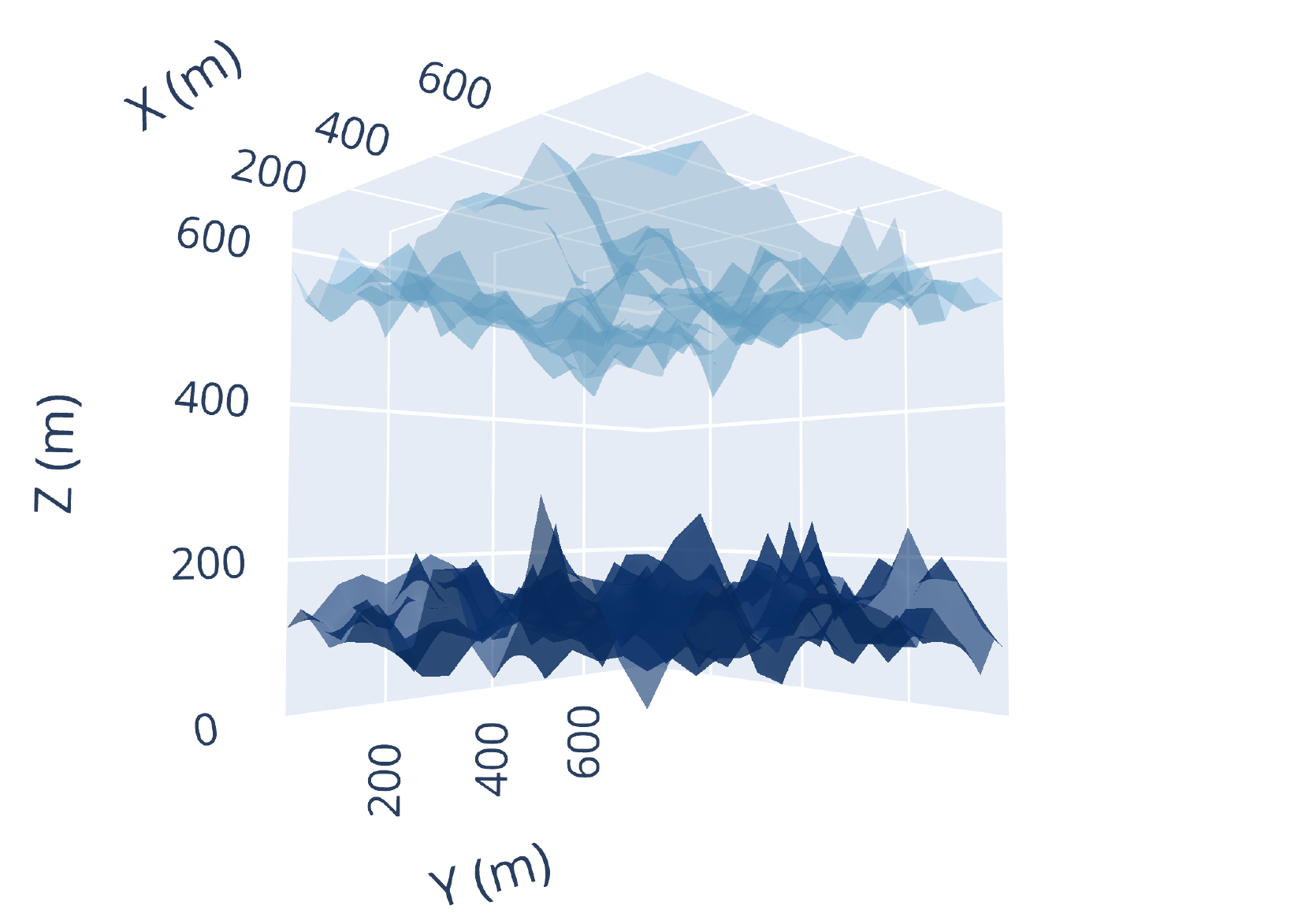}\\
\includegraphics[width=0.25\myfigwidth,trim={165pt 10pt 15pt 0pt},clip]
{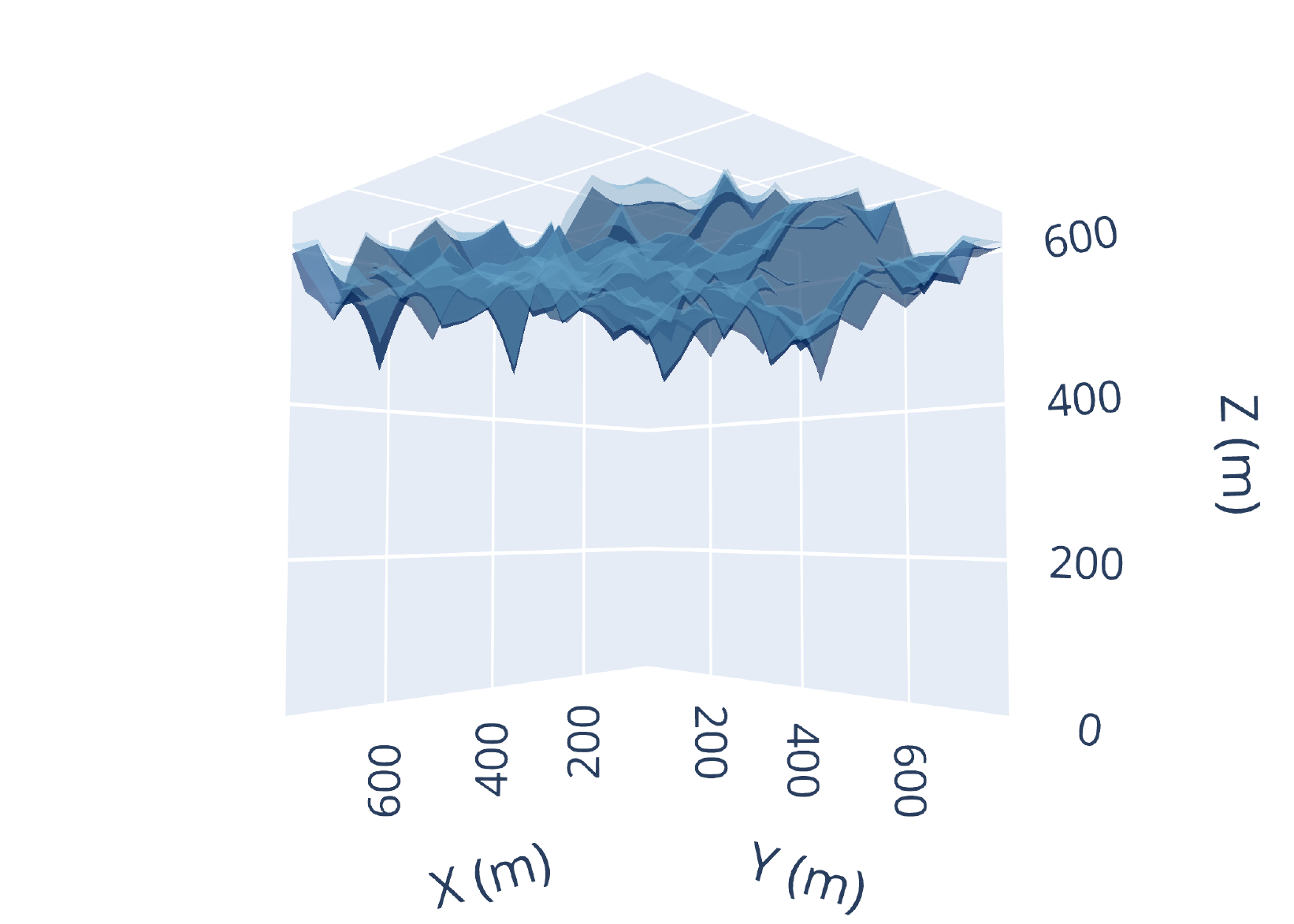}%
\includegraphics[width=0.25\myfigwidth,trim={165pt 10pt 15pt 0pt},clip]
{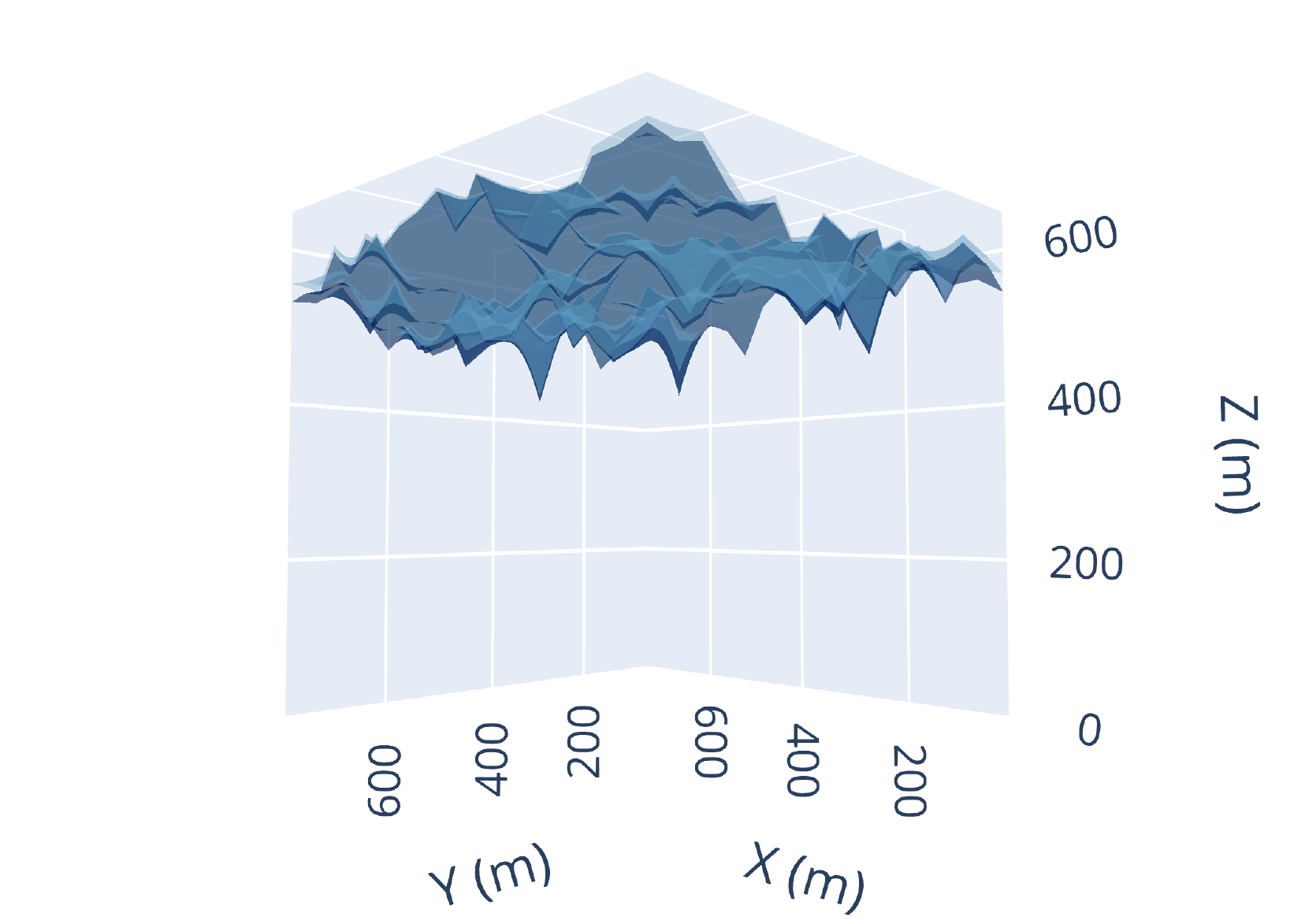}%
\includegraphics[width=0.25\myfigwidth,trim={120pt 10pt 60pt 0pt},clip]
{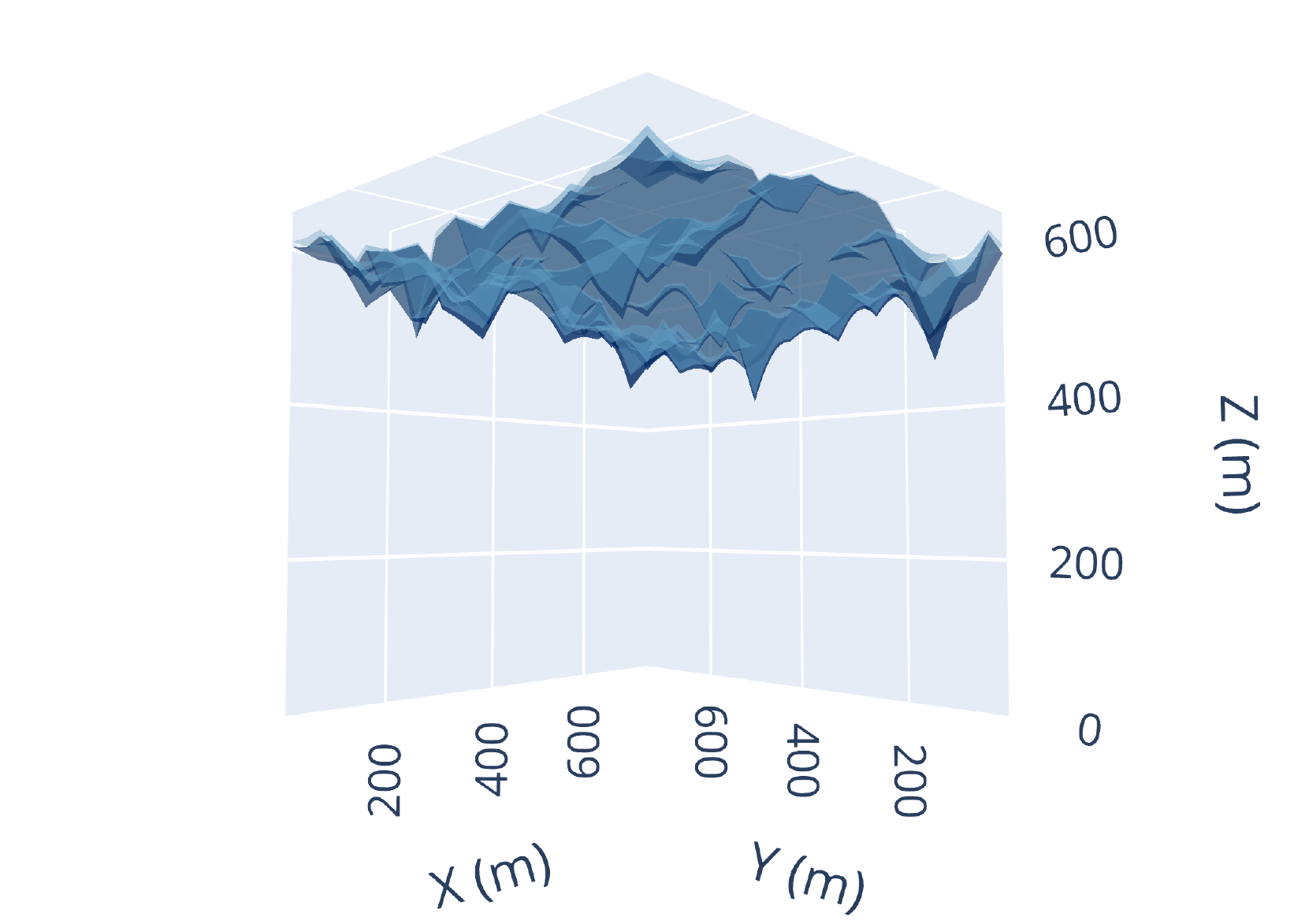}%
\includegraphics[width=0.25\myfigwidth,trim={15pt 10pt 165pt 0pt},clip]
{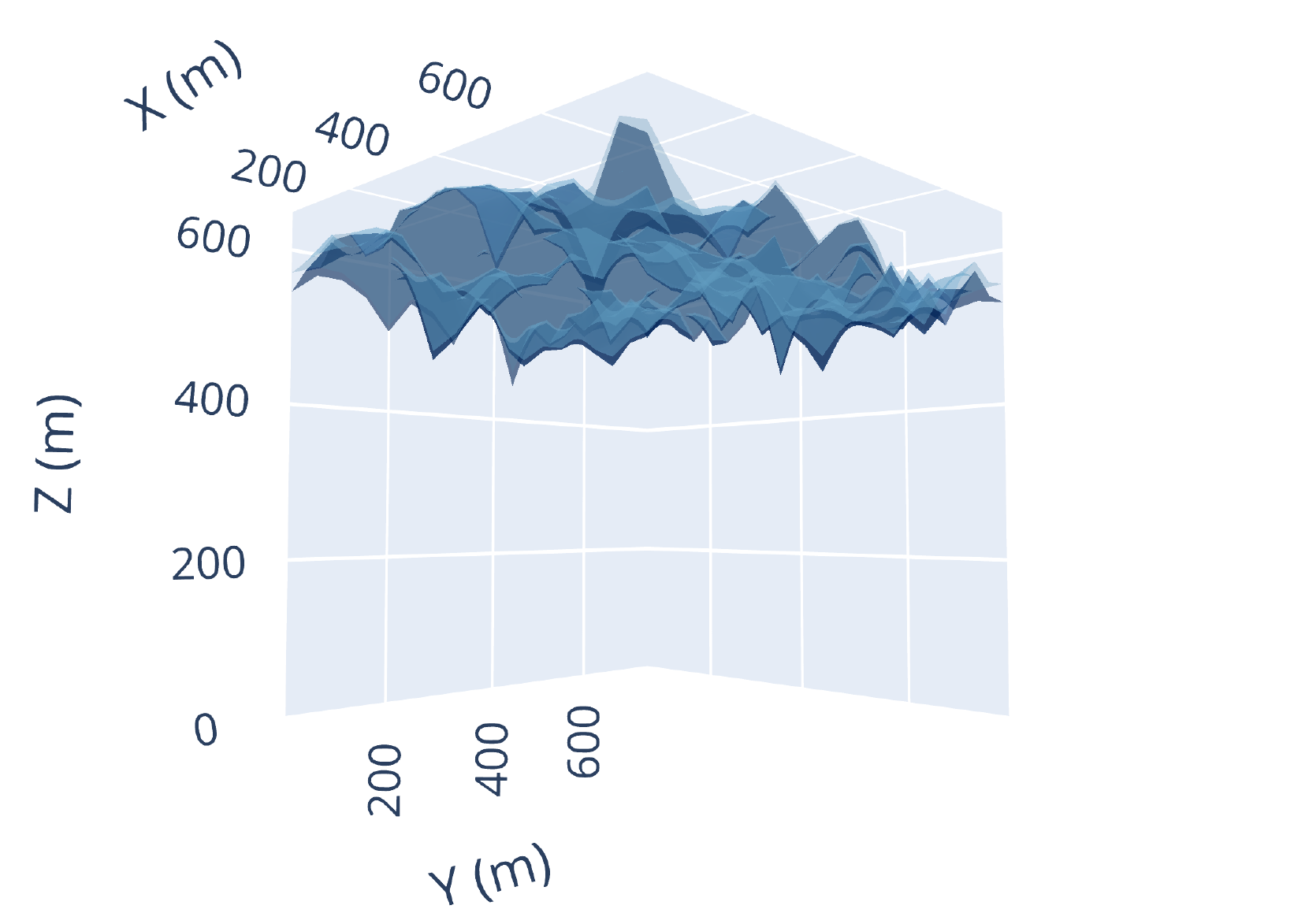}
\caption{Samples from the prior for fixed spatial dependencies parameters 
$r_\ell=0.9999$. Each row represents an independent sample, while each column
corresponds to a $90^\circ$ clockwise rotation about the $z$ axis.}
\label{fig:prior_samples}
\end{myfig}

Here we show a few random layer models drawn from the prior
described in \cref{sec:CAR_prior}. To focus on the more interesting regime 
of high spatial dependence, we fix $r_\ell=0.9999$ for both layers.
\cref{fig:prior_samples} shows three layer models independently sampled from the
prior. We can see that, at this level of spatial dependence, the models look
like stacked wrinkled sheets, aligned with the $(x,y)$ plane. The samples
show varying volumes of air---maximum in the second row, minimum in the last
row---but within each sample the air gap has a fairly constant thickness.
Clearly, these samples do not resemble any kind of block-cave structure, and 
especially not one like the baseline model shown in \cref{fig:true_layers}.
This should not be surprising: the stochastic process described in
\cref{sec:CAR_prior}
uses simple geometric and probabilistic rules to generate layer models.
Nevertheless, as we show in \cref{sec:experiment}, the prior seems to be
flexible enough to allow our inference procedure to recover geometries that
both reproduce the data and resemble the true layer model.

\section{Additional MCMC convergence diagnostics}\label{app:long_mcmc}

\begin{myfig}
\centering
\begin{subfigure}[b]{0.3333\myfigwidth}
\includegraphics[width=\textwidth]{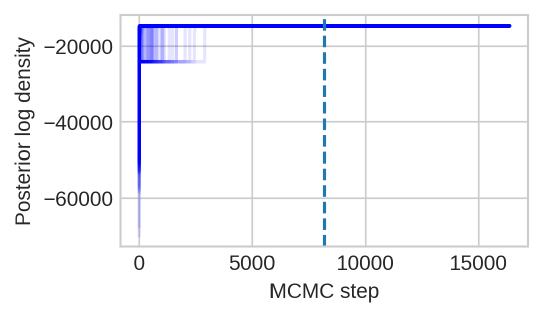}
\vspace{-0.6cm}
\caption{Traces of log density}
\label{fig:long_mcmc_traces_log_density_poisson}
\end{subfigure}%
\begin{subfigure}[b]{0.3333\myfigwidth}
\includegraphics[width=\textwidth]{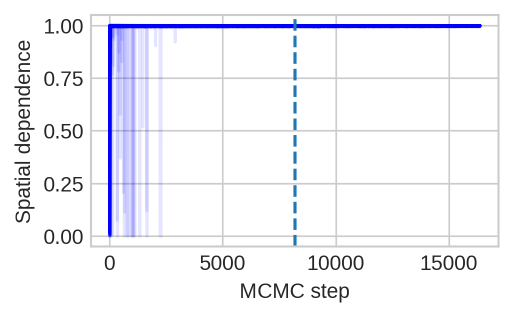}
\vspace{-0.6cm}
\caption{Traces of $r_1$.}
\label{fig:long_mcmc_traces_rho_1_poisson}
\end{subfigure}%
\begin{subfigure}[b]{0.3334\myfigwidth}
\includegraphics[width=\textwidth]{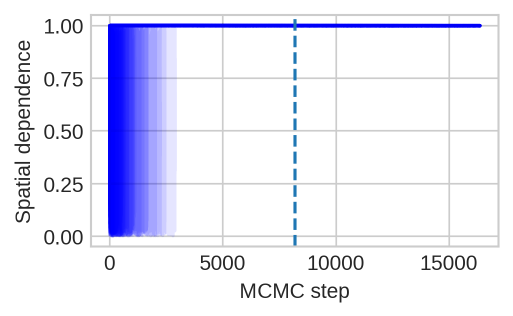}
\vspace{-0.6cm}
\caption{Traces of $r_2$.}
\label{fig:long_mcmc_traces_rho_2_poisson}
\end{subfigure}\\
\begin{subfigure}[b]{0.3333\myfigwidth}
\includegraphics[width=\textwidth]{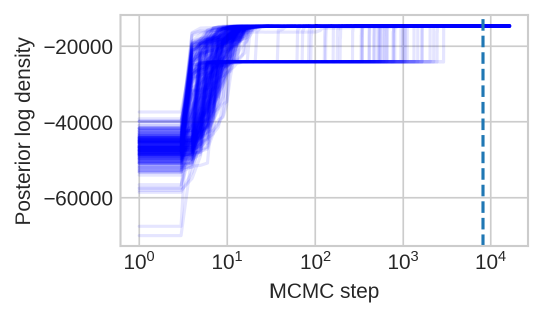}
\vspace{-0.6cm}
\caption{Traces of log density (log)}
\label{fig:long_mcmc_traces_log_density_log_poisson}
\end{subfigure}%
\begin{subfigure}[b]{0.3333\myfigwidth}
\includegraphics[width=\textwidth]{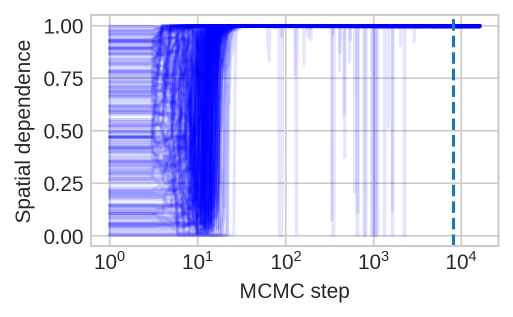}
\vspace{-0.6cm}
\caption{Traces of $r_1$ (log).}
\label{fig:long_mcmc_traces_rho_1_log_poisson}
\end{subfigure}%
\begin{subfigure}[b]{0.3334\myfigwidth}
\includegraphics[width=\textwidth]{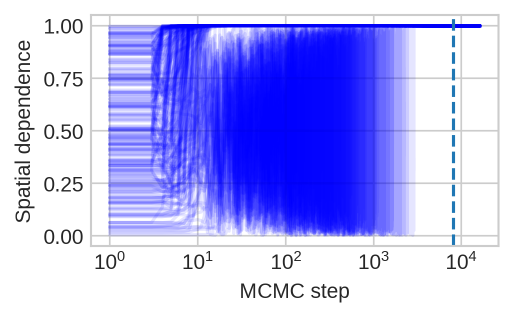}
\vspace{-0.6cm}
\caption{Traces of $r_2$ (log).}
\label{fig:long_mcmc_traces_rho_2_log_poisson}
\end{subfigure}\\
\begin{subfigure}[b]{0.3333\myfigwidth}
\includegraphics[width=\textwidth]{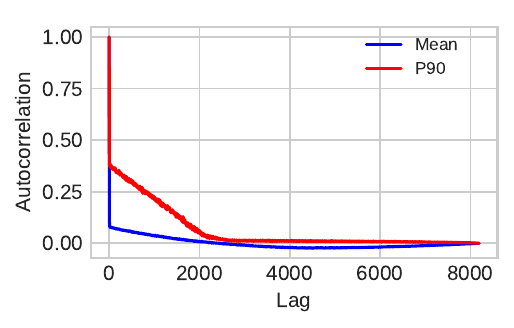}
\vspace{-0.6cm}
\caption{A.C. profile of log density.}
\label{fig:long_mcmc_ac_log_density_poisson}
\end{subfigure}%
\begin{subfigure}[b]{0.3333\myfigwidth}
\includegraphics[width=\textwidth]{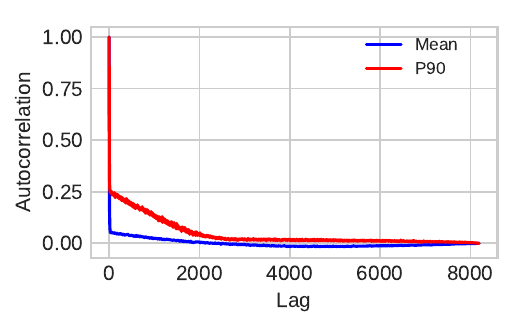}
\vspace{-0.6cm}
\caption{A.C. profile of $r_1$.}
\label{fig:long_mcmc_ac_rho_1_poisson}
\end{subfigure}%
\begin{subfigure}[b]{0.3334\myfigwidth}
\includegraphics[width=\textwidth]{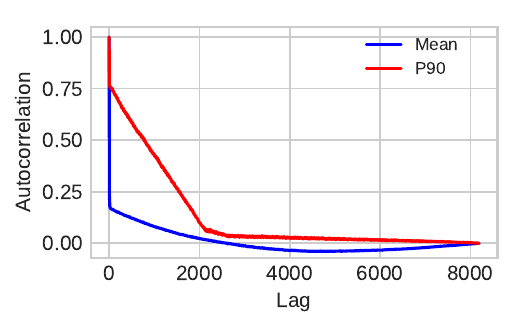}
\vspace{-0.6cm}
\caption{A.C. profile of $r_2$.}
\label{fig:long_mcmc_ac_rho_2_poisson}
\end{subfigure}
\caption{MCMC diagnostics for the posterior log density (un-normalized) and 
spatial dependency parameters. 
\textbf{Top:} traces for all $256$ chains plotted using transparency. 
Vertical dashed lines indicate the last step of the adaptation phase.
\textbf{Middle:} same as above but using $\log_{10}$-transformed $x$-axis to aid
visualization.
\textbf{Bottom:} Mean and $90$-th percentile across chains of the 
post-adaptation auto-correlation (A.C.) profiles.
}
\label{fig:long_mcmc_poisson}
\end{myfig}

Here we present a more classical simulation analysis of the MCMC tools
described in \cref{sec:mcmc}, with the purpose of finding adequate choices for
the number of adaptation and sampling steps required to attain convergence.
This analysis also serves the purpose of corroborating the results obtained via
the many-short-chains framework shown in \cref{sec:experiment}.

We run the same number of $256$ chains recommended in \cref{sec:mcmc}, but with 
all of them initialized at random; i.e., without building super-chains by sharing
initial points. Each chain runs for $2^{13}$ adaptation steps, followed by an
identical number of non-adaptive steps. We store the full trajectory of all
chains, but in order to keep the memory footprint low, we only collect three 
representative statistics: the (un-normalized) posterior log density, and the
two spatial dependence parameters $r_1$ and $r_2$.

\cref{fig:long_mcmc_poisson} shows a summary of the results. The traces at the
top show that chains have concentrated around similar values for the three 
statistics before the first half of the adaptation window finishes. The same 
traces are
plotted again in the middle row with $\log_{10}$-transformed $x$-axis to show
more clearly the behavior during the initial phase. Notice how in this transient
regime many chains are sampling values of $r_1$ and $r_2$ that span the whole 
unit interval, with the latter parameter taking the longest to stabilize. Taken
together, these results suggest that running $2^{13}$ adaptation steps is
enough for the chains to stabilize around the high-posterior-probability region, 
while leaving sufficient time (roughly half of the total adaptation steps) for
the tuning algorithm to optimize the sampler parameters to this specific region.

Finally, the last row of \cref{fig:long_mcmc_poisson} shows a summary of the 
auto-correlation profiles in the post-adaptation phase. To highlight their
variability, we plot the mean and $90$-th percentile across chains.
We see that after $2^{11}$ steps the average auto-correlation is negligible for
all three statistics. However, the $90$-th percentile auto-correlation takes
considerably longer to reach $0$---especially in the case of $r_2$---requiring
almost $2^{13}$ steps. Based on these findings, we believe that running the
super-chains for $2^{13}$ iterations after adaptation should provide a robust
burn-in period to ensure adequate mixing of the chains.

\end{document}